\documentclass[10pt,twocolumn,superscriptaddress,amssymb,amsmath,aps,pra]{revtex4-1}
\usepackage{graphicx}

\begin{document}

\title{Quantized fluctuational electrodynamics for three-dimensional plasmonic structures}
\date{January 30, 2017}
\author{Mikko Partanen}
\affiliation{Engineered Nanosystems group, School of Science, Aalto University, P.O. Box 12200, 00076 Aalto, Finland}
\author{Teppo H\"ayrynen}
\affiliation{DTU Fotonik, Department of Photonics Engineering, Technical University of
Denmark, \O rsteds Plads, Building 343, DK-2800 Kongens Lyngby, Denmark}
\author{Jukka Tulkki}
\affiliation{Engineered Nanosystems group, School of Science, Aalto University, P.O. Box 12200, 00076 Aalto, Finland}
\author{Jani Oksanen}
\affiliation{Engineered Nanosystems group, School of Science, Aalto University, P.O. Box 12200, 00076 Aalto, Finland}

\begin{abstract}
We recently introduced a quantized fluctuational electrodynamics (QFED)
formalism that provides a physically insightful
definition of an effective position-dependent photon-number operator
and the associated ladder operators.
However, this far the formalism has been applicable only for the normal
incidence of the electromagnetic field in planar structures.
In this work, we overcome the main limitation of the one-dimensional QFED
formalism by extending the model to three dimensions, allowing us
to use the QFED method to study, e.g., plasmonic structures.
To demonstrate the benefits of the developed formalism,
we apply it to study the local steady-state photon numbers
and field temperatures in a light-emitting near-surface
InGaN quantum-well structure with a metallic coating
supporting surface plasmons.
\end{abstract}

\maketitle

\section{Introduction}

The quantum optical processes in lossy and lossless material systems have been widely studied
during the last few decades. This has led to advances, e.g., in nanoplasmonics
\cite{Sorger2011,Oulton2009,Huang2013,Sadi2013a},
near-field microscopy \cite{Taubner2006,Hillenbrand2002},
thin-film light-emitting diodes \cite{Nakamura2013,Heikkila2013},
photonic crystals \cite{Russell2003,Akahane2003}, and metamaterials \cite{Tanaka2010,Mattiucci2013}.
For describing spatial field evolution
in resonant structures, one of the most widely used quantization
approaches has been the input-output formalism of the photon creation
and annihilation operators. The formalism was originally developed for describing lossless and dispersionless
dielectrics \cite{Knoll1987} and was later extended for lossy and dispersive media
\cite{Knoll1991,Allen1992,Huttner1992,Barnett1995,Matloob1995,Matloob1996}.
The early studies clearly highlighted that the noise and field operators in nonuniform systems
are position dependent and that the vector potential and electric-field operators
obey the well-known canonical commutation relation
as expected \cite{Barnett1995,Matloob1995}. However,
the canonical commutation relations did not extend to the photon creation and
annihilation operators, which were found to exhibit anomalies in resonant structures
\cite{Ueda1994,Raymer2013,Barnett1996,Aiello2000,Stefano2000}. It was first
concluded that these anomalies have no physical significance.
Although the formalism was later successfully used to study, e.g., amplifying media
and spontaneous decay in left-handed media
\cite{Dung2003,Khanbekyan2005,Raabe2007,Raabe2008},
it was also recently shown that the anomalous commutation
relations should, e.g., lead to the existence of a threshold for
second-harmonic generation inside microcavities \cite{Gauvin2014,Collette2013}.
The anomalous commutation relations have also been found to
prevent a systematic description of local thermal balance
between the field and interacting media \cite{Partanen2014a,Partanen2015a}.

We recently solved the cavity commutation relation anomaly and
photon-number problem by introducing a
quantized fluctuational electrodynamics (QFED) model to describe
photon number and showed that the expectation values of the properly
normalized annihilation and creation operators result in a meaningful
photon-number model and thermal balance conditions
\cite{Partanen2014a,Partanen2014c,Partanen2015a,Partanen2016b}.
This far, our models have been strictly one-dimensional and limited to normal incidence
in planar structures, which has provided an adequate framework for describing the
fundamental  properties of cavity fields. However, considering the associated
transparent description of the photon number and field temperature, it becomes
reasonable to ask how the description can be expanded to more complex systems involving, e.g.,
plasmons that have been of great topical interest
\cite{Barnes2003,Okamoto2004,Yeh2008,Bonnand2006a,Bonnand2006b,Tanaka2010,Pitarke2007}
and whose description could benefit
from the new methodology clearly separating the local density of states (LDOS)
and the photon number.
Here we therefore present a generalized QFED model to account
for fully three-dimensional propagation as well as the associated spectral expansion
for planar structures. We also demonstrate the
benefits of the formalism by applying it to study the local steady-state
field properties and plasmonic interactions in a light-emitting near-surface
InGaN quantum-well (QW) structure with a metallic coating
supporting surface plasmons (SPs).

This paper is organized as follows: The theory of the QFED method
is presented in Sec.~\ref{sec:theory}. 
As a background for QFED, we
first review the general three-dimensional
noise operator formalism 
and the use of Green's functions to obtain the solutions
of the electromagnetic (EM) fields. This is followed by
a presentation of the new contribution to the theory:
the properly normalized position- and frequency-dependent photon ladder operators,
the related photon-number presentation, and the
generalized forms of the densities of states
in the QFED method. After introducing the ladder
and number operators, we also briefly
review how the operators can be used to present
the associated models for the field fluctuations,
Poynting vector, and absorption and emission operators. Note that
the expectation values of these macroscopic field quantities 
are equivalent to the values obtained by using
the conventional fluctuational electrodynamics.
In Sec.~\ref{sec:results}, we demonstrate the applicability and study 
the physical implications of the introduced QFED method in an example
InGaN QW geometry.

\section{\label{sec:theory}Quantized fluctuational electrodynamics method}

In this section, we outline the derivation of the three-dimensional QFED
theory. Detailed derivations are given in the appendixes.
We start by introducing the fundamental equations of the conventional fluctuational
electrodynamics theory and its quantization in Sec.~\ref{sec:noiseoperatorformalism}
and the solution of fields using the dyadic Green's functions in Sec.~\ref{sec:green}.
Then, in Sec.~\ref{sec:photonnumbers}, we present the properly normalized photon
ladder operators and the related photon-number and density-of-states concepts
that expand the classical and previously used quantized versions of fluctuational 
electrodynamics to enable an unambiguous photon-level description
of the three-dimensional system. In Secs.~\ref{sec:poynting} and \ref{sec:balance},
we focus on calculating the Poynting vector
operator and the thermal balance predicted by the quantized theory
using the newly established operators.

\subsection{\label{sec:noiseoperatorformalism}Noise operator formalism}
Maxwell's equations describe electric and magnetic fields generated by currents
and charges in matter. They relate the electric field strength $\mathbf{E}$,
the magnetic field strength $\mathbf{H}$,
the electric flux density $\mathbf{D}$, and
the magnetic flux density $\mathbf{B}$ to the free electric charge density
$\rho_\mathrm{f}$ and the free electric current density $\mathbf{J}_\mathrm{f}$.
In the frequency domain, Maxwell's equations are written for positive frequencies as \cite{Partanen2016b}
\begin{align}
 \nabla\cdot\mathbf{D} &=\rho_\mathrm{f},\label{eq:maxwell1}\\
 \nabla\cdot\mathbf{B} &=0,\label{eq:maxwell2}\\
 \nabla\times\mathbf{E} &=i\omega\mathbf{B}=i\omega\mu_0(\mu\mathbf{H}+\delta\mathbf{M}),\label{eq:maxwell3}\\
 \nabla\times\mathbf{H} &=\mathbf{J}_\mathrm{f}-i\omega\mathbf{D}=\mathbf{J}_\mathrm{f}-i\omega(\varepsilon_0\varepsilon\mathbf{E}+\delta\mathbf{P}).\label{eq:maxwell4}
\end{align}
Here we have additionally used the constitutive relations $\mathbf{D}=\varepsilon_0\varepsilon\mathbf{E}+\delta\mathbf{P}$
and $\mathbf{B}=\mu_0(\mu\mathbf{H}+\delta\mathbf{M})$,
where $\varepsilon_0$ and $\mu_0$
are the permittivity and permeability of vacuum,
$\varepsilon=\varepsilon_\mathrm{r}+i\varepsilon_\mathrm{i}$
and $\mu=\mu_\mathrm{r}+i\mu_\mathrm{i}$
are the relative permittivity and permeability of the medium
with real and imaginary
parts denoted by subscripts $\mathrm{r}$ and $\mathrm{i}$,
and the polarization and magnetization fields
$\delta\mathbf{P}$ and $\delta\mathbf{M}$ denote the polarization
and magnetization that are not linearly proportional to the respective field
strengths \cite{Sipe1987}.
In the context of the fluctuational electrodynamics and the present work,
$\delta\mathbf{P}$ and $\delta\mathbf{M}$ describe the
small thermal fluctuations of the linear polarization and 
magnetization fields \cite{Partanen2016b}.
For the remainder of this work, the
current density of free charges $\mathbf{J}_\mathrm{f}$ is also included in
the electric permittivity for notational simplicity.

From Maxwell's equations in Eqs.~\eqref{eq:maxwell1}--\eqref{eq:maxwell4}
it follows that the electric field obeys the well-known equation \cite{Partanen2016b}
\begin{align}
 \nabla\times\Big(\frac{\nabla\times\mathbf{E}}{\mu_0\mu}\Big)-\omega^2\varepsilon_0\varepsilon\mathbf{E} &=i\omega\mathbf{J}_\mathrm{e}
 -\nabla\times\Big(\frac{\mathbf{J}_\mathrm{m}}{\mu_0\mu}\Big),\label{eq:HelmholtzE1}
\end{align}
where the terms $\mathbf{J}_\mathrm{e}=-i\omega\delta\mathbf{P}$
and $\mathbf{J}_\mathrm{m}=-i\omega\mu_0\delta\mathbf{M}$ represent
the polarization and magnetization currents that act as field
sources in the noise operator theory \cite{Partanen2016b} and in the classical fluctuational
electrodynamics \cite{Narayanaswamy2014,Polimeridis2015}.
The electric term $\mathbf{J}_\mathrm{e}$ includes contributions from both the electric currents
due to free charges and polarization terms associated with
dipole currents and thermal dipole fluctuations. For the magnetic current term $\mathbf{J}_\mathrm{m}$,
the only contribution arises from the magnetic dipoles.
Note that, after solving the electric field from Eq.~\eqref{eq:HelmholtzE1},
the calculation of the magnetic field is straightforward using
Faraday's law in Eq.~\eqref{eq:maxwell3}.

In the previously known noise operator framework, we use the canonical quantization of the above classical
equations where the classical field
quantities are replaced by corresponding quantum operators \cite{Dung2003,Matloob1995,Partanen2014a,Partanen2016b}.
The electric and magnetic noise current operators $\hat{\mathbf{J}}_\mathrm{e}^+(\mathbf{r},\omega)$
and $\hat{\mathbf{J}}_\mathrm{m}^+(\mathbf{r},\omega)$ are written in terms
of bosonic source field operators $\hat f_\mathrm{e}(\mathbf{r},\omega)$ and
$\hat f_\mathrm{m}(\mathbf{r},\omega)$ as
$\hat{\mathbf{J}}_\mathrm{e}^+(\mathbf{r},\omega)=\sum_\alpha j_\mathrm{0,e}(\mathbf{r},\omega)\hat{\mathbf{e}}_\alpha\hat f_\mathrm{e}(\mathbf{r},\omega)$ and
$\hat{\mathbf{J}}_\mathrm{m}^+(\mathbf{r},\omega)=\sum_\alpha j_\mathrm{0,m}(\mathbf{r},\omega)\hat{\mathbf{e}}_\alpha\hat f_\mathrm{m}(\mathbf{r},\omega)$,
where $\hat{\mathbf{e}}_\alpha$ are unit vectors
for the three coordinate directions $\alpha\in\{x,y,z\}$
\cite{Partanen2016b,Partanen2014a}.
The operators $\hat f_\mathrm{e}(\mathbf{r},\omega)$ and
$\hat f_\mathrm{m}(\mathbf{r},\omega)$ obey
the canonical commutation relation
$[\hat f_{j}(\mathbf{r},\omega),\hat f_k^\dag(\mathbf{r}',\omega')]=\delta_{jk}\delta(\mathbf{r}-\mathbf{r}')\delta(\omega-\omega')$,
with $j,k\in\{\mathrm{e,m}\}$.
The normalization factors $j_\mathrm{0,e}(\mathbf{r},\omega)$ and $j_\mathrm{0,m}(\mathbf{r},\omega)$ have
been determined to be
$j_\mathrm{0,e}(\mathbf{r},\omega)=\sqrt{4\pi\hbar\omega^2\varepsilon_0\varepsilon_\mathrm{i}(\mathbf{r},\omega)}$
\cite{Partanen2014a,Khanbekyan2005} and
$j_\mathrm{0,m}(\mathbf{r},\omega)=\sqrt{4\pi\hbar\omega^2\mu_0\mu_\mathrm{i}(\mathbf{r},\omega)}$
\cite{Partanen2016b,Dung2003}.

\subsection{\label{sec:green}Green's functions}

In order to write the solution of Eq.~\eqref{eq:HelmholtzE1}
in a general form, we apply the conventional dyadic Green's function formalism \cite{Eckhardt1984,Paulus2000},
where the field solutions are written as
\begin{align}
 \hat{\mathbf{E}}^+(\mathbf{r},\omega)
 &=i\omega\mu_0\int\overset{\text{\tiny$\leftrightarrow$}}{\mathbf{G}}_\mathrm{ee}(\mathbf{r},\omega,\mathbf{r}')\cdot\hat{\mathbf{J}}_\mathrm{e}^+(\mathbf{r}',\omega)d^3r'\nonumber\\
 &\hspace{0.5cm}+k_0\int\overset{\text{\tiny$\leftrightarrow$}}{\mathbf{G}}_\mathrm{em}(\mathbf{r},\omega,\mathbf{r}')\cdot\hat{\mathbf{J}}_\mathrm{m}^+(\mathbf{r}',\omega)d^3r',
 \label{eq:efield}
\end{align}
\begin{align}
 \hat{\mathbf{H}}^+(\mathbf{r},\omega)
 &=k_0\int\overset{\text{\tiny$\leftrightarrow$}}{\mathbf{G}}_\mathrm{me}(\mathbf{r},\omega,\mathbf{r}')\cdot\hat{\mathbf{J}}_\mathrm{e}^+(\mathbf{r}',\omega)d^3r'\nonumber\\
 &\hspace{0.5cm}+i\omega\varepsilon_0\int\overset{\text{\tiny$\leftrightarrow$}}{\mathbf{G}}_\mathrm{mm}(\mathbf{r},\omega,\mathbf{r}')\cdot\hat{\mathbf{J}}_\mathrm{m}^+(\mathbf{r}',\omega)d^3r'.
 \label{eq:hfield}
\end{align}
Here $\overset{\text{\tiny$\leftrightarrow$}}{\mathbf{G}}_\mathrm{ee}(\mathbf{r},\omega,\mathbf{r}')$
is the electric Green's function,
$\overset{\text{\tiny$\leftrightarrow$}}{\mathbf{G}}_\mathrm{mm}(\mathbf{r},\omega,\mathbf{r}')$
is the magnetic Green's function, and
$\overset{\text{\tiny$\leftrightarrow$}}{\mathbf{G}}_\mathrm{em}(\mathbf{r},\omega,\mathbf{r}')$ and
$\overset{\text{\tiny$\leftrightarrow$}}{\mathbf{G}}_\mathrm{me}(\mathbf{r},\omega,\mathbf{r}')$
are the exchange Green's functions.
For completeness, the relations between these Green's functions
are shown in Appendix \ref{apx:greengeneral}, and the Green's
functions are explicitly presented for stratified media in Appendix \ref{apx:greenstratified}.

\subsection{\label{sec:photonnumbers}Photon numbers and densities of states}

In analogy with the one-dimensional QFED formalism \cite{Partanen2016b},
we can define the position- and frequency-dependent effective photon ladder operators
$\hat a_j(\mathbf{r},\omega)$, which obey the canonical commutation relation
$[\hat a_j(\mathbf{r},\omega),\hat a_j^\dag(\mathbf{r},\omega)]=\delta(\omega-\omega')$,
for the electric, magnetic, and total EM fields, $j\in\{\mathrm{e,m,tot}\}$.
These operators and the corresponding effective photon-number expectation values
$\langle\hat n_\mathrm{j}(\mathbf{r},\omega')\rangle$ are given by
\begin{align}
 \hat a_j(\mathbf{r},\omega)&=\frac{1}{\sqrt{\int\rho_{\mathrm{NL},j}(\mathbf{r},\omega,\mathbf{r}')d^3r'}}\nonumber\\
 &\hspace{0.5cm}\times\int\Big[\sqrt{\rho_{\mathrm{NL},j,\mathrm{e}}(\mathbf{r},\omega,\mathbf{r}')}\hat f_\mathrm{e}(\mathbf{r}',\omega)\nonumber\\
 &\hspace{0.5cm}+\sqrt{\rho_{\mathrm{NL},j,\mathrm{m}}(\mathbf{r},\omega,\mathbf{r}')}\hat f_\mathrm{m}(\mathbf{r}',\omega)\Big]d^3r',
 \label{eq:photonladder}
\end{align}
\begin{equation}
 \langle\hat n_{j}(\mathbf{r},\omega)\rangle=\frac{\int\rho_{\mathrm{NL},j}(\mathbf{r},\omega,\mathbf{r}')\langle\hat\eta(\mathbf{r}',\omega)\rangle d^3r'}{\int\rho_{\mathrm{NL},j}(\mathbf{r},\omega,\mathbf{r}')d^3r'},
 \label{eq:photonnumbers}
\end{equation}
where $\langle\hat\eta(\mathbf{r}',\omega)\rangle$ is the source-field photon-number
expectation value related to the bosonic noise operators as
$\langle\hat\eta(\mathbf{r},\omega)\rangle
=\int\langle\hat f_\mathrm{e}^\dag(\mathbf{r},\omega)\hat f_\mathrm{e}(\mathbf{r}',\omega')\rangle d^3r'd\omega'
=\int\langle\hat f_\mathrm{m}^\dag(\mathbf{r},\omega)\hat f_\mathrm{m}(\mathbf{r}',\omega')\rangle d^3r'd\omega'$
and $\rho_{\mathrm{NL},j}(\mathbf{r},\omega,\mathbf{r}')$
are the nonlocal densities of states (NLDOSs) for the electric,
magnetic, and total EM fields, written as
\begin{align}
 &\rho_\mathrm{NL,e}(\mathbf{r},\omega,\mathbf{r}')\nonumber\\
 &=\frac{2\omega^3}{\pi c^4}\Big(\varepsilon_\mathrm{i}(\mathbf{r}',\omega)
\mathrm{Tr}[\overset{\text{\tiny$\leftrightarrow$}}{\mathbf{G}}_\mathrm{ee}(\mathbf{r},\omega,\mathbf{r}')\cdot
 \overset{\text{\tiny$\leftrightarrow$}}{\mathbf{G}}_\mathrm{ee}^\dag(\mathbf{r},\omega,\mathbf{r}')]\nonumber\\
&\hspace{0.5cm}+\mu_\mathrm{i}(\mathbf{r}',\omega)
\mathrm{Tr}[\overset{\text{\tiny$\leftrightarrow$}}{\mathbf{G}}_\mathrm{em}(\mathbf{r},\omega,\mathbf{r}')\cdot
 \overset{\text{\tiny$\leftrightarrow$}}{\mathbf{G}}_\mathrm{em}^\dag(\mathbf{r},\omega,\mathbf{r}')]\Big)
\label{eq:enldos},\\
 &\rho_\mathrm{NL,m}(\mathbf{r},\omega,\mathbf{r}')\nonumber\\
 &=\frac{2\omega^3}{\pi c^4}\Big(\varepsilon_\mathrm{i}(\mathbf{r}',\omega)
\mathrm{Tr}[\overset{\text{\tiny$\leftrightarrow$}}{\mathbf{G}}_\mathrm{me}(\mathbf{r},\omega,\mathbf{r}')\cdot
 \overset{\text{\tiny$\leftrightarrow$}}{\mathbf{G}}_\mathrm{me}^\dag(\mathbf{r},\omega,\mathbf{r}')]\nonumber\\
&\hspace{0.5cm}+\mu_\mathrm{i}(\mathbf{r}',\omega)
\mathrm{Tr}[\overset{\text{\tiny$\leftrightarrow$}}{\mathbf{G}}_\mathrm{mm}(\mathbf{r},\omega,\mathbf{r}')\cdot
 \overset{\text{\tiny$\leftrightarrow$}}{\mathbf{G}}_\mathrm{mm}^\dag(\mathbf{r},\omega,\mathbf{r}')]\Big)
\label{eq:hnldos},\\
 &\rho_\mathrm{NL,tot}(\mathbf{r},\omega,\mathbf{r}')\nonumber\\
 &=\frac{|\varepsilon(\mathbf{r},\omega)|}{2}\rho_\mathrm{NL,e}(\mathbf{r},\omega,\mathbf{r}')
 +\frac{|\mu(\mathbf{r},\omega)|}{2}\rho_\mathrm{NL,m}(\mathbf{r},\omega,\mathbf{r}').
 \label{eq:unldos}
\end{align}
The NLDOS components $\rho_{\mathrm{NL},j,\mathrm{e}}(\mathbf{r},\omega,\mathbf{r}')$ and
$\rho_{\mathrm{NL},j,\mathrm{m}}(\mathbf{r},\omega,\mathbf{r}')$, with $j\in\{\mathrm{e,m}\}$,
in Eq.~\eqref{eq:photonladder}
denote, respectively, the first and the second terms of Eqs.~\eqref{eq:enldos} and \eqref{eq:hnldos}.
The total NLDOS terms $\rho_{\mathrm{NL,tot,e}}(\mathbf{r},\omega,\mathbf{r}')$ and
$\rho_{\mathrm{NL,j,m}}(\mathbf{r},\omega,\mathbf{r}')$ are calculated by using Eq.~\eqref{eq:unldos}
with the corresponding terms in the electric and magnetic NLDOSs.

Note that the expressions for the photon ladder operators and the photon numbers
in Eqs.~\eqref{eq:photonladder} and \eqref{eq:photonnumbers}
are the same as the expression in the one-dimensional formalism \cite{Partanen2016b},
but the NLDOSs are different.
The derivation of these NLDOSs is presented in Appendix \ref{apx:densities},
and for general stratified media, the densities of states are presented in Appendix \ref{apx:stratifieddos}.
The LDOSs $\rho_j(\mathbf{r},\omega)$ are given in terms of the NLDOSs by
\begin{equation}
 \rho_j(\mathbf{r},\omega)=\int\rho_{\mathrm{NL},j}(\mathbf{r},\omega,\mathbf{r}')d^3r'.
 \label{eq:ldos}
\end{equation}

It is well-known that, in vacuum, the imaginary parts of the traces of the dyadic Green's
functions give the electric and magnetic LDOSs
$\rho_\mathrm{e}(\mathbf{r},\omega)$ and $\rho_\mathrm{m}(\mathbf{r},\omega)$ as \cite{Joulain2003,Joulain2005}
\begin{align}
 \rho_{j}(\mathbf{r},\omega) &=\frac{2\omega}{\pi c^2}\mathrm{Im}\{\mathrm{Tr}[\overset{\text{\tiny$\leftrightarrow$}}{\mathbf{G}}_{jj}(\mathbf{r},\omega,\mathbf{r})]\},
 \label{eq:traceldos}
\end{align}
where $j\in\{\mathrm{e,m}\}$.
A similar relation also applies for the normal components of the Fourier-transformed
quantities in layered media \cite{Partanen2014c,Partanen2016b}, and typically, also, the spatially
resolved form in Eq.~\eqref{eq:traceldos} is expected to be valid inside lossy media.
However, in lossy media, these LDOSs are generally known to become infinite
due to the contribution of evanescent waves \cite{Joulain2003,Joulain2005}.

In terms of the photon-number expectation values in Eq.~\eqref{eq:photonnumbers} and
the LDOSs in Eq.~\eqref{eq:ldos},
the spectral electric and magnetic field fluctuations and the energy density are given by \cite{Partanen2014c}
\begin{align}
 \langle\hat{E}(\mathbf{r},t)^2\rangle_\omega & =\frac{\hbar\omega}{\varepsilon_0}\rho_\mathrm{e}(\mathbf{r},\omega)\Big(\langle\hat n_\mathrm{e}(\mathbf{r},\omega)\rangle+\frac{1}{2}\Big)\label{eq:efluct},\\[8pt]
 \langle\hat{H}(\mathbf{r},t)^2\rangle_\omega & =\frac{\hbar\omega}{\mu_0}\rho_\mathrm{m}(\mathbf{r},\omega)\Big(\langle\hat n_\mathrm{m}(\mathbf{r},\omega)\rangle+\frac{1}{2}\Big)\label{eq:bfluct},\\[8pt]
 \langle\hat u(\mathbf{r},t)\rangle_\omega & = \hbar\omega\rho_\mathrm{tot}(\mathbf{r},\omega)\Big(\langle\hat n_\mathrm{tot}(\mathbf{r},\omega)\rangle+\frac{1}{2}\Big)\label{eq:edensity}.
\end{align}
Here the subscript $\omega$ denotes the contribution of $\omega$
to the total quantities which are obtained as integrals over positive frequencies.

\subsection{\label{sec:poynting}Quantized Poynting vector operator}

To conform with our earlier works and to enable describing energy flow in detail,
we also find the three-dimensional generalized expression for the Poynting vector.
For an optical mode, the
quantum optical Poynting vector is defined as a normal-ordered operator
in terms of the positive- and negative-frequency parts of the electric 
and magnetic field operators as
$\hat{\mathbf{S}}(\mathbf{r},t)=:\!\hat{\mathbf{E}}(\mathbf{r},t)\times\hat{\mathbf{H}}(\mathbf{r},t)\!:=\hat{\mathbf{E}}^-(\mathbf{r},t)\times\hat{\mathbf{H}}^+(\mathbf{r},t)-\hat{\mathbf{H}}^-(\mathbf{r},t)\times\hat{\mathbf{E}}^+(\mathbf{r},t)$
\cite{Loudon2000}. As detailed in Appendix \ref{apx:densities} and
in analogy with the one-dimensional QFED formalism \cite{Partanen2014a,Partanen2016b}, we
obtain the Poynting-vector-related interference density of states (IFDOS) as
\begin{align}
 &\boldsymbol{\rho}_\mathrm{IF}(\mathbf{r},\omega,\mathbf{r}')\nonumber\\
 &=\frac{2\omega^3n_\mathrm{r}(\mathbf{r},\omega)}{\pi c^4}\nonumber\\
 &\hspace{0.5cm}\times\Big(\mu_\mathrm{i}(\mathbf{r}',\omega)
 \mathrm{Im}\Big[\mathrm{Tr}[\overset{\text{\tiny$\leftrightarrow$}}{\mathbf{G}}_\mathrm{mm}(\mathbf{r},\omega,\mathbf{r}')\times
 \overset{\text{\tiny$\leftrightarrow$}}{\mathbf{G}}_\mathrm{em}^\dag(\mathbf{r},\omega,\mathbf{r}')]\Big]\nonumber\\
 &\hspace{0.5cm}-\varepsilon_\mathrm{i}(\mathbf{r}',\omega)
 \mathrm{Im}\Big[\mathrm{Tr}[\overset{\text{\tiny$\leftrightarrow$}}{\mathbf{G}}_\mathrm{ee}(\mathbf{r},\omega,\mathbf{r}')\times
 \overset{\text{\tiny$\leftrightarrow$}}{\mathbf{G}}_\mathrm{me}^\dag(\mathbf{r},\omega,\mathbf{r}')]\Big]\Big),
\end{align}
where $n_\mathrm{r}(\mathbf{r},\omega)$ is the real part of the refractive index and
we have used the short-hand notation
$\mathrm{Tr}[\overset{\text{\tiny$\leftrightarrow$}}{\mathbf{G}}_{1}(\mathbf{r},\omega,\mathbf{r}')\times
 \overset{\text{\tiny$\leftrightarrow$}}{\mathbf{G}}_{2}^\dag(\mathbf{r},\omega,\mathbf{r}')]
 =\sum_\alpha[\overset{\text{\tiny$\leftrightarrow$}}{\mathbf{G}}_{1}(\mathbf{r},\omega,\mathbf{r}')\cdot\hat{\mathbf{e}}_\alpha]\times
 [\overset{\text{\tiny$\leftrightarrow$}}{\mathbf{G}}_{2}(\mathbf{r},\omega,\mathbf{r}')\cdot\hat{\mathbf{e}}_\alpha]^\dag$,
which is a vector, in contrast to the conventional trace of a matrix.
Using the IFDOS, the Poynting vector is given by
\begin{equation}
 \langle\hat{\mathbf{S}}(\mathbf{r},t)\rangle_\omega =
 \hbar\omega v(\mathbf{r},\omega)\int\boldsymbol{\rho}_\mathrm{IF}(\mathbf{r},\omega,\mathbf{r}')\langle\hat\eta(\mathbf{r}',\omega)\rangle d^3r',
 \label{eq:poynting}
\end{equation}
where $v(\mathbf{r},\omega)=c/n_\mathrm{r}(\mathbf{r},\omega)$ is
the propagation velocity of the field in the direction of the wave vector.
The integral of the IFDOS with respect to $\mathbf{r}'$ is always zero, i.e.,
$\int\boldsymbol{\rho}_\mathrm{IF}(\mathbf{r},\omega,\mathbf{r}')d^3r'=0$, which is required by
the fact that, in a medium in thermal equilibrium, there is no net energy flow.
For stratified media, the IFDOS is presented in Appendix \ref{apx:stratifieddos}.

\subsection{\label{sec:balance}Field-matter interaction operators and thermal balance}

A particularly insightful view of the effective photon numbers
in the QFED framework is provided by their connection
to local thermal balance between the field and matter \cite{Partanen2014a}.
First, we define the normal-ordered emission and absorption operators
$\hat Q_\mathrm{em}(\mathbf{r},t)$ and $\hat Q_\mathrm{abs}(\mathbf{r},t)$ as
\begin{align}
 \hat Q_\mathrm{em}(\mathbf{r},t)\!
 &=\!-\!:\!\hat{\mathbf{J}}_\mathrm{e}(\mathbf{r},t)\cdot\hat{\mathbf{E}}(\mathbf{r},t)\!:
 \!-\!:\!\hat{\mathbf{J}}_\mathrm{m}(\mathbf{r},t)\cdot\hat{\mathbf{H}}(\mathbf{r},t)\!:,
 \label{eq:emissionop}\\
 \hat Q_\mathrm{abs}(\mathbf{r},t)\!
 &=:\!\hat{\mathbf{J}}_\mathrm{e,abs}(\mathbf{r},t)\cdot\hat{\mathbf{E}}(\mathbf{r},t)\!:
 \!+\!:\!\hat{\mathbf{J}}_\mathrm{m,abs}(\mathbf{r},t)\cdot\hat{\mathbf{H}}(\mathbf{r},t)\!:,
 \label{eq:absorptionop}
\end{align}
where the electric and magnetic absorption current operators $\hat{\mathbf{J}}_\mathrm{e,abs}(\mathbf{r},t)$ and
$\hat{\mathbf{J}}_\mathrm{m,abs}(\mathbf{r},t)$ are written
in the spectral domain as
$\hat{\mathbf{J}}_\mathrm{e,abs}^+(\mathbf{r},\omega)=-i\omega\varepsilon_0\chi_\mathrm{e}(\mathbf{r},\omega)\hat{\mathbf{E}}^+(\mathbf{r},\omega)$
and 
$\hat{\mathbf{J}}_\mathrm{m,abs}^+(\mathbf{r},\omega)=-i\omega\mu_0\chi_\mathrm{m}(\mathbf{r},\omega)\hat{\mathbf{H}}^+(\mathbf{r},\omega)$,
where $\chi_\mathrm{e}(\mathbf{r},\omega)=\varepsilon(\mathbf{r},\omega)-1$ and $\chi_\mathrm{m}(\mathbf{r},\omega)=\mu(\mathbf{r},\omega)-1$
are the electric and magnetic susceptibilities of the medium.

The net emission operator $\hat Q(\mathbf{r},t)=\hat Q_\mathrm{em}(\mathbf{r},t)-\hat Q_\mathrm{abs}(\mathbf{r},t)$,
which describes the energy transfer between the EM field and the local medium, is given by
\begin{align}
 \hat Q(\mathbf{r},t)
 =&:\!\hat{\mathbf{J}}_\mathrm{e,tot}(\mathbf{r},t)\cdot\hat{\mathbf{E}}(\mathbf{r},t)\!:+:\!\hat{\mathbf{J}}_\mathrm{m,tot}(\mathbf{r},t)\cdot\hat{\mathbf{H}}(\mathbf{r},t)\!:,
 \label{eq:netemissionop}
\end{align}
where $\hat{\mathbf{J}}_\mathrm{e,tot}(\mathbf{r},t)=\hat{\mathbf{J}}_\mathrm{e}(\mathbf{r},t)+\hat{\mathbf{J}}_\mathrm{e,abs}(\mathbf{r},t)$
and
$\hat{\mathbf{J}}_\mathrm{m,tot}(\mathbf{r},t)=\hat{\mathbf{J}}_\mathrm{m}(\mathbf{r},t)+\hat{\mathbf{J}}_\mathrm{m,abs}(\mathbf{r},t)$
correspond to the classical total current densities, 
which are sums of free and bound current densities.
The spectral component of the expectation value of the net emission operator
in Eq.~\eqref{eq:netemissionop} can be written in terms of the LDOSs
and the electric- and magnetic-field photon numbers as
\begin{align}
 &\langle\hat Q(\mathbf{r},t)\rangle_\omega\nonumber\\
 & =\hbar\omega^2\varepsilon_\mathrm{i}(\mathbf{r},\omega)\rho_\mathrm{e}(\mathbf{r},\omega)[\langle\hat\eta(\mathbf{r},\omega)\rangle-\langle\hat n_\mathrm{e}(\mathbf{r},\omega)\rangle]\nonumber\\
&\hspace{0.5cm}+\hbar\omega^2\mu_\mathrm{i}(\mathbf{r},\omega)\rho_\mathrm{m}(\mathbf{r},\omega)[\langle\hat\eta(\mathbf{r},\omega)\rangle-\langle\hat n_\mathrm{m}(\mathbf{r},\omega)\rangle].
 \label{eq:divP}
\end{align}
This shows that local thermal balance [$\langle\hat Q(\mathbf{r},t)\rangle_\omega=0$]
is generally reached when the source-field photon numbers coincide with the field photon numbers
as defined in Eq.~\eqref{eq:divP}. In addition, the net emission operator satisfies
$\langle\hat Q(\mathbf{r},t)\rangle_\omega=\nabla\cdot\langle\hat{\mathbf{S}}(\mathbf{r},t)\rangle_\omega$.
In resonant systems where the energy exchange is dominated by a narrow frequency band,
the condition $\langle\hat Q(\mathbf{r},t)\rangle_\omega=0$ can be used
to  approximately determine the steady-state temperature of a weakly interacting resonant
particle \cite{Bohren1998}.

\section{\label{sec:results}Results}

\begin{figure}[b]
\includegraphics[width=0.48\textwidth]{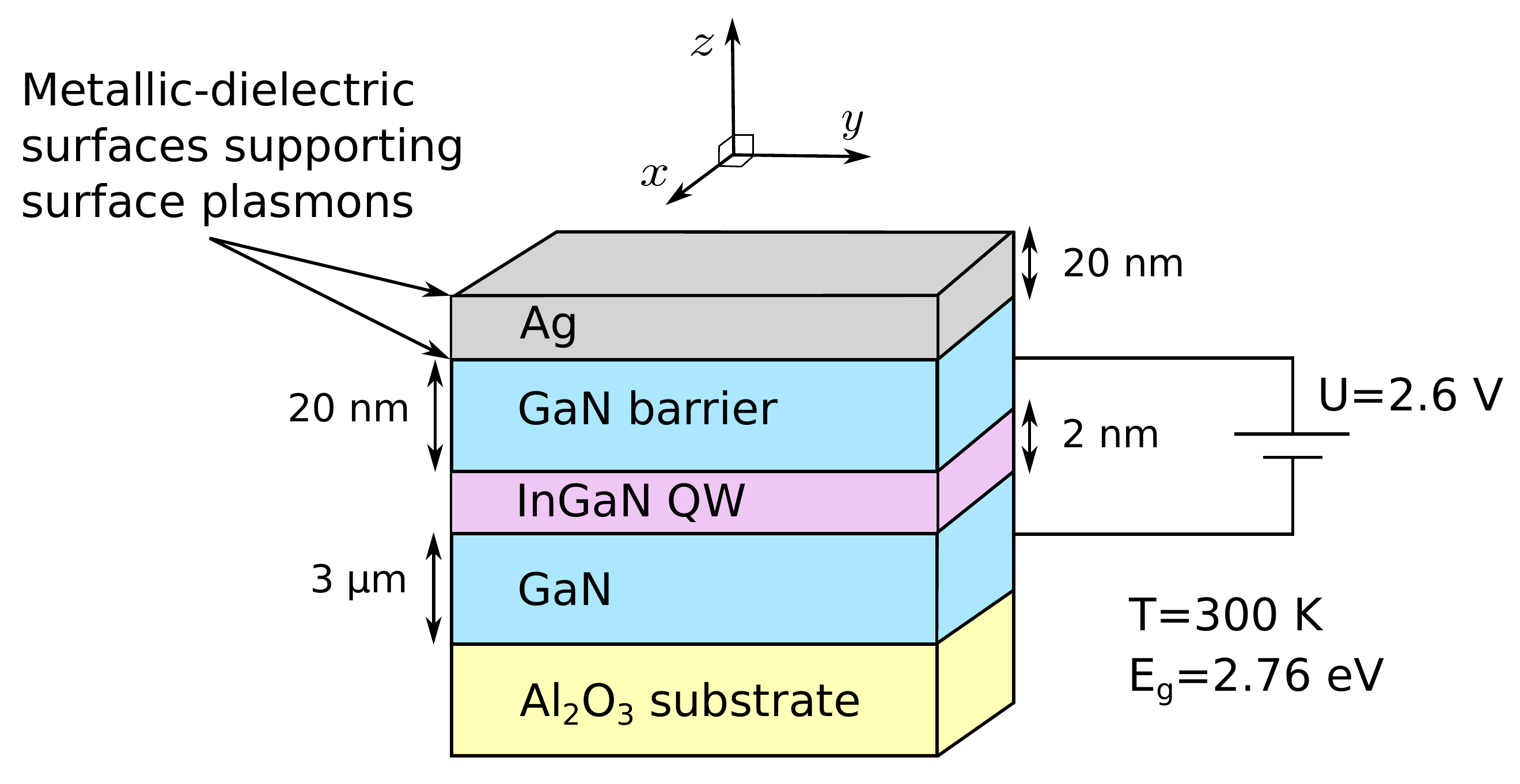}
\caption{\label{fig:structure}(Color online) The studied structure formed by
a Ag/GaN/In$_{0.15}$Ga$_{0.85}$N/GaN/Al$_2$O$_3$ heterostructure. The background temperature is
$T=300$ K, the band gap of the light emitting In$_{0.15}$Ga$_{0.85}$N QW is $E_\mathnormal{g}=2.76$ eV, and the
QW excitation corresponds to an applied voltage of $U=2.6$ V.  Note that the figure is not to scale.}
\end{figure}

\begin{figure*}
\includegraphics[width=\textwidth]{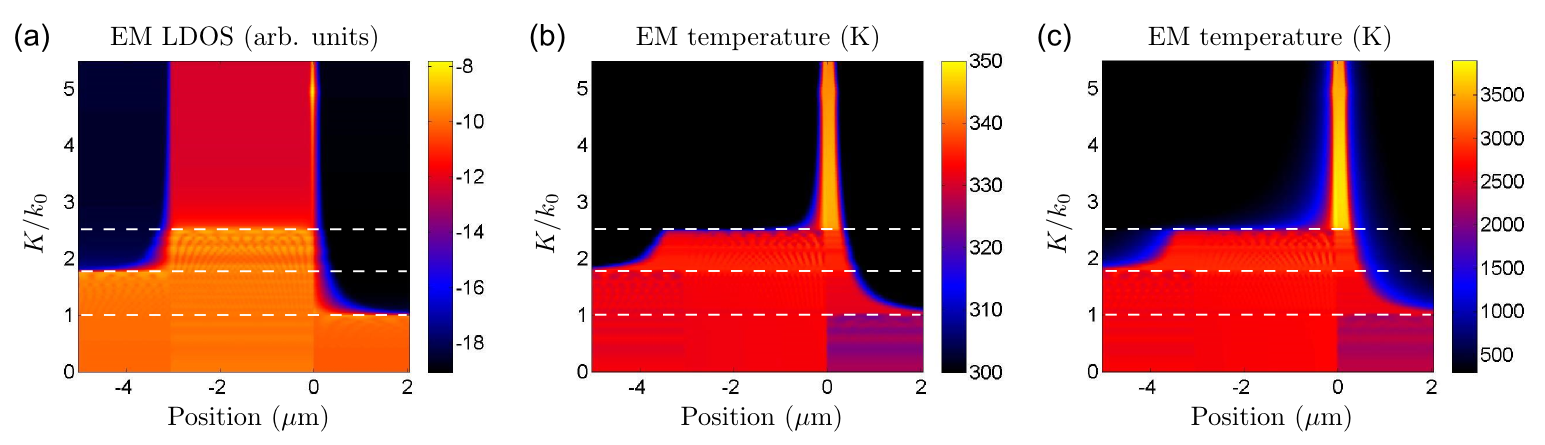}
\caption{\label{fig:pos}(Color online) (a) The base-10 logarithm of the
total EM LDOS, (b) the effective temperature of
the total EM field in the case of a thermally excited QW, and
(c) the effective-field temperature in the case of an electrically or optically excited QW
corresponding to the bias voltage $U=2.6$ V
for photon energy $\hbar\omega=E_\mathnormal{g}+k_\mathrm{B}T=2.786$ eV
as a function of position and the in-plane component of the wave vector.
The position $z=0$ is fixed to the Ag/air interface. The white dashed lines
represent the light cones of GaN, sapphire, and air.}
\end{figure*}

We apply the QFED formalism presented in Sec.~\ref{sec:theory} to the study of an example
plasmonic multilayer structure, which has recently been of experimental and theoretical interest
\cite{Homeyer2013,Sadi2013b}.
In contrast to the previous QFED models, the generalized model can, among other oblique-angle problems,
describe the optical properties of plasmonic devices.
Here we study the contribution of the evanescent SP modes to the
position-dependent LDOSs and the effective-field temperatures in
the vicinity of a light-emitting Ag/GaN/In$_{0.15}$Ga$_{0.85}$N/GaN/Al$_2$O$_3$ multilayer
structure illustrated in Fig.~\ref{fig:structure}.
The 2-nm In$_{0.15}$Ga$_{0.85}$N QW
has a band gap of 2.76 eV ($\lambda=450$ nm), and it
acts as the emitter layer.
It is deposited 20 nm below the 20-nm silver layer which
supports SP modes.
The refractive indices of GaN and InN are taken from Refs.~\citenum{Leung1998,Barker1973,Ambacher1996,Djurisic1999,Trainor1974,Tansley1986},
and the refractive index of In$_{0.15}$Ga$_{0.85}$N is deduced by using Vegard's law;
the refractive index of silver is calculated by using the Drude model
plasma frequency $\omega_\mathrm{p}=9.04$ eV/$\hbar$ and damping frequency $\omega_\tau=0.02125$ eV/$\hbar$
taken from Ref.~\citenum{Zeman1987}, and
the refractive index of sapphire is taken from Ref.~\citenum{Malitson1962}.
For example, in the case of the photon energy $\hbar\omega=2.76$ eV
corresponding to the QW band gap, the refractive indices
of air, silver, GaN, In$_{0.15}$Ga$_{0.85}$N, and sapphire are 1.00, $0.013+3.119i$, $2.51+0.0029i$,
$2.51+0.094i$, and 1.78, respectively.

The background temperature of the materials is $T=300$ K.
We compare the emission of the structure in two cases: (1) the QW is thermally excited
to temperature $T_\mathrm{ex}=350$ K, and (2) the QW
is electrically or optically excited to a state corresponding to direct excitation
by a $U=2.6$ V voltage source.
In the first case, the QW source-field photon-number
expectation value is modeled using the Bose-Einstein distribution
$\langle\hat\eta_\text{\tiny QW}\rangle=1/(e^{\hbar\omega/(k_\mathrm{B}T_\mathrm{ex})}-1)$.
In other words, we apply the local thermal equilibrium (LTE) approximation.
The LTE approximation is justified when the gradients in the temperature
are expected to be small compared to a material-dependent current-current correlation
length scale, which is of the order of atomic scale or the phonon mean free path \cite{Polimeridis2015}.
In the second case, the source-field photon number of the QW is modeled using
$\langle\hat\eta_\text{\tiny QW}\rangle=1/(e^{(\hbar\omega-eU)/(k_\mathrm{B}T)}-1)$
for photon energies above the band gap $\hbar\omega\ge E_\mathnormal{g}$
and the background value $\langle\hat\eta_\text{\tiny BG}\rangle=1/(e^{\hbar\omega/(k_\mathrm{B}T)}-1)$
for photon energies below the band gap $\hbar\omega\le E_\mathnormal{g}$
corresponding to the interactions with the free carriers.
For example, in the case of the photon energy $\hbar\omega=E_\mathnormal{g}+k_\mathrm{B}T=2.786$ eV,
the source-field photon number of the electrically or optically
excited QW is $\langle\hat\eta_\text{\tiny QW}\rangle=7.51\times 10^{-4}$,
which is very large in comparison with the photon number of the thermal 300 K background
$\langle\hat\eta_\text{\tiny BG}\rangle=1.57\times 10^{-47}$.
As the photon-number expectation values are relatively small and depend strongly on the frequency,
it is convenient to illustrate the results by using
the effective field temperature that is defined in terms of
the photon-number expectation value as
$T_\mathrm{eff}(z,K,\omega)=\hbar\omega/\{(k_\mathrm{B}\ln[1+1/\langle\hat n(z,K,\omega)\rangle]\}$
\cite{Partanen2014c,Partanen2014b}. The corresponding effective-source-field temperature of the
electrically or optically excited QW ranges from 5175 K (compare with $\sim$6000 K of solar radiation on earth) to 625 K as the photon energy
ranges from 2.76 to 5 eV.

Figure \ref{fig:pos}(a) shows the base-10 logarithm of the
total EM LDOS for photon energy $\hbar\omega=E_\mathnormal{g}+k_\mathrm{B}T=2.786$ eV as a function of position
and the in-plane component of the wave vector. The sapphire substrate lies
on the left and air on the right. The light cones for sapphire,
GaN, and air are defined by the in-plane wave vector component values $K<nk_0$,
where $n$ is the real part of the refractive index of the respective material.
The light cones of the different material layers are clearly visible in the 
figure. Due to the evanescent fields, the LDOSs are slightly elevated also
beyond the material interfaces. One can also see the
very large LDOS associated with the GaN/Ag SP resonance
near the position $z=0$ and $K/k_0=5.0$.
The less visible air/Ag SP resonance is
near the position $z=0$ and $K/k_0=1.0$.
The GaN guided modes and the associated interference patterns
can be seen between the GaN light cone and
the sapphire light cone with $1.78<K/k_0<2.51$.

\begin{figure*}
\includegraphics[width=\textwidth]{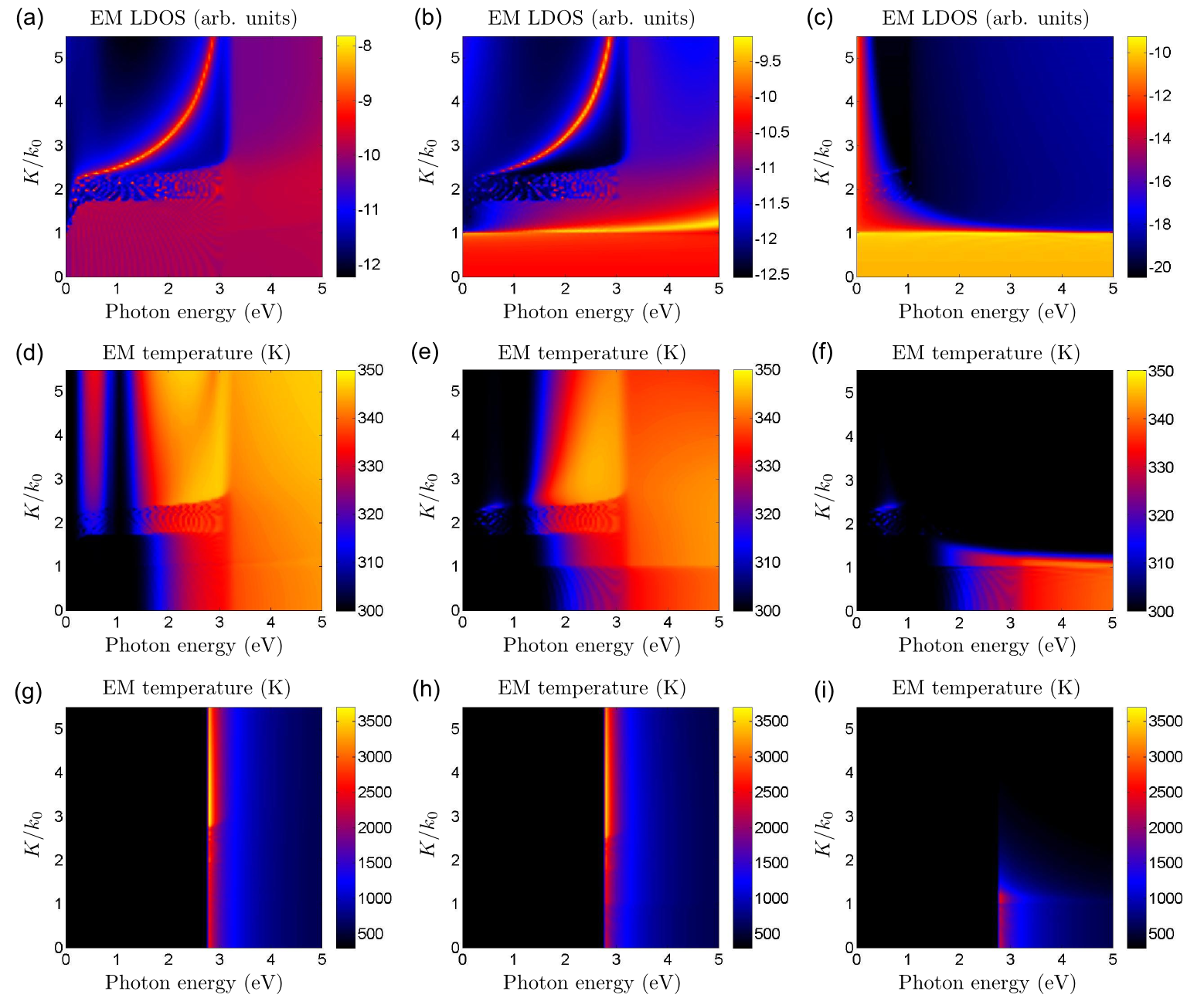}
\caption{\label{fig:ene}(Color online) The base-10 logarithm of the
total EM LDOS as a function of photon energy and
the in-plane component of the wave vector (a) in the QW, (b) in air at 1 nm above the surface,
and (c) in air at 1 $\mu$m above the surface. (d), (e), and (f) 
The effective temperature of the total EM field at the corresponding positions
in the case of a thermally excited QW.
(g), (h), and (i) The effective-field temperature at the corresponding
positions in the case of
an electrically or optically excited QW
corresponding to the bias voltage $U=2.6$ V.}
\end{figure*}

Figure \ref{fig:pos}(b) shows the effective-field temperature of the total EM field
corresponding to the LDOS in Fig.~\ref{fig:pos}(a) in the case of a thermally excited QW.
For the narrow In$_{0.15}$Ga$_{0.85}$N layer located slightly left
from the position $z=0$ $\mu$m,
the source-field temperature is 350 K; for other material layers it is 300 K.
It can be seen that the light cones
of each material are visible also in the effective-field temperature.
The evanescent fields near the material interfaces are even more pronounced
when compared to the LDOS in Fig.~\ref{fig:pos}(a).
At high values of $K$, the effective-field temperatures approach the source-field
temperature in each layer, whereas for K within the light cones of air
and sapphire, the effective temperature is reduced due to the strong
coupling to the semi-infinite air and sapphire layers.
Figure \ref{fig:pos}(c) shows the corresponding effective-field temperature of the total EM field
in the case of an electrically or optically excited QW.
The figure clearly resembles the case of thermal excitation in Fig.~\ref{fig:pos}(b),
but the values of the effective-field temperature are significantly higher,
as expected.

Figure \ref{fig:ene}(a) presents the base-10 logarithm of the total
EM LDOS in the QW as a function of the photon energy and the in-plane component of
the wave vector. The figure clearly shows the GaN/Ag SP resonance, as well as the
GaN guided modes corresponding to the Fabry-P\'erot resonances of the cavity.
At photon energy slightly above 3 eV, the GaN becomes absorptive, and therefore,
there are no resonances visible above this energy.
If, instead of the total EM LDOS, we were to plot the LDOS
parts corresponding to the TE and TM polarizations,
then the SP modes would be visible only in the TM case,
as previously discussed, e.g., in Ref.~\cite{Sadi2013b}.
Otherwise, the LDOSs of the TE and TM polarizations are qualitatively
very similar. Figure \ref{fig:ene}(b) shows the corresponding
base-10 logarithm of the total EM LDOS in air at 1 nm above the structure.
In addition to the resonances visible in Fig.~\ref{fig:ene}(a),
in Fig.~\ref{fig:ene}(b), one can also see the 
Ag/air SP mode just above the light cone of air.
Figure \ref{fig:ene}(c) presents the base-10 logarithm
of the EM LDOS in air at 1 $\mu$m above the structure.
One can clearly see that there is only a small contribution
of the evanescent fields remaining, especially at high frequencies,
and the only significant contribution to the
EM LDOS arises from the propagating modes in
the light cone of air.

Figure \ref{fig:ene}(d) shows the effective-field temperature of the
total EM field in the middle of the QW
as a function of energy and $K/k_0$ for
the case of a thermally excited QW. The effective temperature
is essentially above the background temperature of 300 K when
the imaginary part of the refractive index of the InGaN QW significantly
deviates from zero either due to band-to-band or other
absorption and emission mechanisms. At low frequencies, the thin InGaN becomes
nearly transparent, and therefore, the effective temperature reaches the background
temperature. The emissivity peak near photon
energy $\hbar\omega=0.5$ eV follows from the peak in the
infrared absorption coefficient of the QW \cite{Tansley1986}.
The corresponding effective-field temperature in the case of an
electrically or optically excited QW is shown in Fig.~\ref{fig:ene}(g).
The emission begins at the photon energy corresponding to the
band gap, where the resulting
effective-field temperature also obtains its highest values,
as expected. At high energies well above the band gap,
the effective-field temperature again reaches the source-field
temperature of 300 K.
The effective temperature of the field
generally increases as the optical confinement of the mode increases:
In the case of an electrically or optically excited QW and
photon energy $\hbar\omega=E_\mathnormal{g}+k_\mathrm{B}T=2.786$ eV,
the  modes extending into the light cone of air have $T_\mathrm{eff}\approx 2200$ K,
whereas the modes bound in the light cone of GaN reach $T_\mathrm{eff}\approx 2700$ K,
while the evanescent InGaN modes reach values as high as $T_\mathrm{eff}\approx 3500$ K.
For all these cases, however, $T_\mathrm{eff}$ remains well below the
source-field temperature of the QW due to the losses caused by the
surrounding lossy materials.

Figure \ref{fig:ene}(e) presents the effective-field temperature of the
total EM field in air at 1 nm above the structure,
corresponding to the LDOS in Fig.~\ref{fig:ene}(b)
in the case of a thermally excited QW. The values
of the effective-field temperature are somewhat lower than
the values of the effective-field temperature in the QW
in Fig.~\ref{fig:ene}(d). This is mainly due to the attenuation
related to the increased distance to the excited QW.
The effective-field temperature in Fig.~\ref{fig:ene}(e),
however, resembles the effective-field temperature in the QW.
Also, the effective-field temperatures in the case of an electrically or optically
excited QW at the two positions presented in Figs.~\ref{fig:ene}(g) and Fig.~\ref{fig:ene}(h)
are quite similar.
In the case of a thermally excited QW at low frequencies,
the infrared emission of the QW is not visible in air
as the silver layer between air and the QW
becomes very lossy at low frequencies.

Figures \ref{fig:ene}(f) and \ref{fig:ene}(i) show the effective-field
temperatures of the total EM field in air at 1 $\mu$m above the structure
corresponding to the LDOS in Fig.~\ref{fig:ene}(c)
in the cases of thermally and electrically or optically excited QWs.
The contribution of the evanescent fields is reduced
as in the case of the EM LDOS. Due to the longer distance
to the structure and reflections at the interfaces, the values of the effective-field temperatures
are also consequently lower compared to the values of the effective-field
temperatures in the QW in Figs.~\ref{fig:ene}(d) and \ref{fig:ene}(g).

\section{\label{sec:conclusions}Conclusions}

We have developed a three-dimensional QFED method
to describe the photon-number quantization and thermal balance
in general lossy and lossless geometries. By appropriately defining
the photon ladder operators and the densities of states,
we were able to present the ladder operators and the photon-number expectation
values using formulas that are equivalent to the forms
previously obtained by using a one-dimensional formalism.
The resulting generalized QFED method allows studying, e.g., plasmonic
structures and defining an effective-field temperature that
realistically describes the excitation of the optical field.

To demonstrate the applicability and physical
implications of the presented QFED method, we have used
the model to study the energy and position dependencies of the
EM LDOSs and effective-field temperatures
in a light-emitting InGaN QW structure, which has recently
been of experimental and theoretical interest.
The results show that the developed method is well suited
for analyzing the emission of electrically, optically,
or thermally excited QWs. The effective temperatures were
studied both as a function of position and as a function
of photon energy. Electrical and optical excitations of the
QW produce high effective-field temperatures, whose energy
spectrum is quite narrow, whereas the effective-field
temperature of a thermally excited QW
has a significantly broader emission spectrum, as expected.

In addition to providing further insight into the classical
fluctuational electrodynamics theory,
the QFED method enables interesting further studies as it
bridges the classical propagating wave
picture of the EM field and the fluctuational
electrodynamics, which is widely used to model near-field
effects. Therefore, we expect that using the
QFED, method one could, for instance, find a radiative
transfer equation that allows describing interference effects,
thus widening the applicability
of the conventional radiative transfer equation beyond its
main limitation in describing interference effects.
This would make it possible to use the radiative transfer
equation to describe also near-field effects in resonant structures.

\begin{acknowledgments}
This work has been funded in part by the Academy of Finland and the Aalto Energy Efficiency Research Programme.
\end{acknowledgments}

\appendix

\section{\label{apx:greengeneral}Green's functions}

Here we briefly review the known relations between the electric,
magnetic, and exchange Green's functions.
We first define the electric Green's function
$\overset{\text{\tiny$\leftrightarrow$}}{\mathbf{G}}_\mathrm{ee}(\mathbf{r},\omega,\mathbf{r}')$
that satisfies \cite{Paulus2000}
\begin{align}
 &\nabla_\mathbf{r}\times\Big(\frac{\nabla_\mathbf{r}\times\overset{\text{\tiny$\leftrightarrow$}}{\mathbf{G}}_\mathrm{ee}(\mathbf{r},\omega,\mathbf{r}')}{\mu(\mathbf{r},\omega)}\Big)-k_0^2\varepsilon(\mathbf{r},\omega)\overset{\text{\tiny$\leftrightarrow$}}{\mathbf{G}}_\mathrm{ee}(\mathbf{r},\omega,\mathbf{r}')\nonumber\\
 &=\overset{\text{\tiny$\leftrightarrow$}}{\mathbf{I}}\delta(\mathbf{r}-\mathbf{r}'),
 \label{eq:greenee}
\end{align}
where $\overset{\text{\tiny$\leftrightarrow$}}{\mathbf{I}}$ is the unit dyadic
and $k_0=\omega/c$ is the wavenumber in vacuum with the vacuum velocity of light $c$.
The subscript $\mathbf{r}$ in $\nabla_\mathbf{r}$ highlights that the differentiation
is here performed with respect to $\mathbf{r}$ instead of $\mathbf{r}'$.
The solution of Eq.~\eqref{eq:HelmholtzE1} is then written in terms of the electric Green's function
$\overset{\text{\tiny$\leftrightarrow$}}{\mathbf{G}}_\mathrm{ee}(\mathbf{r},\omega,\mathbf{r}')$
by integrating the product of the Green's function and the source terms over all the source points $\mathbf{r}'$ as
\begin{align}
 \hat{\mathbf{E}}^+(\mathbf{r},\omega)
 &= \mu_0\int\overset{\text{\tiny$\leftrightarrow$}}{\mathbf{G}}_\mathrm{ee}(\mathbf{r},\omega,\mathbf{r}')\cdot\Big[i\omega\hat{\mathbf{J}}_\mathrm{e}^+(\mathbf{r}',\omega)\nonumber\\
 &\hspace{0.5cm}-\nabla_{\mathbf{r}'}\times\Big(\frac{\hat{\mathbf{J}}_\mathrm{m}^+(\mathbf{r}',\omega)}{\mu_0\mu(\mathbf{r}',\omega)}\Big)\Big]d^3r'\nonumber\\
 &=i\omega\mu_0\int\overset{\text{\tiny$\leftrightarrow$}}{\mathbf{G}}_\mathrm{ee}(\mathbf{r},\omega,\mathbf{r}')\cdot\hat{\mathbf{J}}_\mathrm{e}^+(\mathbf{r}',\omega)d^3r'\nonumber\\
 &\hspace{0.5cm}+k_0\int\overset{\text{\tiny$\leftrightarrow$}}{\mathbf{G}}_\mathrm{em}(\mathbf{r},\omega,\mathbf{r}')\cdot\hat{\mathbf{J}}_\mathrm{m}^+(\mathbf{r}',\omega)d^3r',
 \label{eq:efield2}
\end{align}
where the subscript $\mathbf{r}'$ in $\nabla_{\mathbf{r}'}$ inside the integral indicates that the differentiation
is performed with respect to the source point $\mathbf{r}'$.
In the case of the second term, we have applied the Stokes' theorem resulting in the integration by parts formula
$\int_V\mathbf{G}^\alpha\cdot(\nabla_{\mathbf{r}'}\times\mathbf{J})d^3r'=\int_V(\nabla_{\mathbf{r}'}\times\mathbf{G}^\alpha)\cdot\mathbf{J}d^3r'-\int_{\partial V}(\mathbf{G}^\alpha\times\mathbf{J})\cdot d\mathbf{S}'$
separately for each row vector $\mathbf{G}^\alpha$ of the matrix representation of
$\overset{\text{\tiny$\leftrightarrow$}}{\mathbf{G}}_\mathrm{ee}(\mathbf{r},\omega,\mathbf{r}')$
with the boundary condition that the Green's functions go to zero when the separation
between the source point $\mathbf{r}'$ and the field point $\mathbf{r}$ tends to infinity.
Using the short hand notation
$\overset{\text{\tiny$\leftrightarrow$}}{\mathbf{G}}_\mathrm{ee}(\mathbf{r},\omega,\mathbf{r}')\times\nabla_{\mathbf{r}'}
=-[\nabla_{\mathbf{r}'}\times\mathbf{G}^1,\nabla_{\mathbf{r}'}\times\mathbf{G}^2,\nabla_{\mathbf{r}'}\times\mathbf{G}^3]^T$,
where $T$ denotes transpose, we then define the
exchange Green's function $\overset{\text{\tiny$\leftrightarrow$}}{\mathbf{G}}_\mathrm{em}(\mathbf{r},\omega,\mathbf{r}')$ as
\begin{equation}
 \overset{\text{\tiny$\leftrightarrow$}}{\mathbf{G}}_\mathrm{em}(\mathbf{r},\omega,\mathbf{r}')
 =\frac{\overset{\text{\tiny$\leftrightarrow$}}{\mathbf{G}}_\mathrm{ee}(\mathbf{r},\omega,\mathbf{r}')\times\nabla_{\mathbf{r}'}}{k_0\mu(\mathbf{r}',\omega)}.
 \label{eq:greenem}
\end{equation}

Solving for the magnetic field by using Faraday's law in Eq.~\eqref{eq:maxwell3}
and substituting the electric field operator in terms of the Green's functions in Eq.~\eqref{eq:efield2} give
\begin{align}
 \hat{\mathbf{H}}^+(\mathbf{r},\omega) &=\frac{1}{i\omega\mu_0\mu(\mathbf{r},\omega)}\Big(\hat{\mathbf{J}}_\mathrm{m}^+(\mathbf{r},\omega)+\nabla_\mathbf{r}\times\hat{\mathbf{E}}^+(\mathbf{r},\omega)\Big)\nonumber\\
 &=k_0\int\frac{\nabla_\mathbf{r}\times\overset{\text{\tiny$\leftrightarrow$}}{\mathbf{G}}_\mathrm{ee}(\mathbf{r},\omega,\mathbf{r}')}{k_0\mu(\mathbf{r},\omega)}\cdot\hat{\mathbf{J}}_\mathrm{e}^+(\mathbf{r}',\omega)d^3r'\nonumber\\
 &\hspace{0.5cm}-\frac{ik_0^2}{\omega\mu_0}\int\Big[\frac{\nabla_\mathbf{r}\times\overset{\text{\tiny$\leftrightarrow$}}{\mathbf{G}}_\mathrm{em}(\mathbf{r},\omega,\mathbf{r}')}{k_0\mu(\mathbf{r},\omega)}\nonumber\\
 &\hspace{0.5cm}+\overset{\text{\tiny$\leftrightarrow$}}{\mathbf{I}}\frac{\delta(\mathbf{r}-\mathbf{r}')}{k_0^2\mu(\mathbf{r},\omega)}\Big]\cdot\hat{\mathbf{J}}_\mathrm{m}^+(\mathbf{r}',\omega)d^3r'\nonumber\\
 &=k_0\int\overset{\text{\tiny$\leftrightarrow$}}{\mathbf{G}}_\mathrm{me}(\mathbf{r},\omega,\mathbf{r}')\cdot\hat{\mathbf{J}}_\mathrm{e}^+(\mathbf{r}',\omega)d^3r'\nonumber\\
 &\hspace{0.5cm}+i\omega\varepsilon_0\int\overset{\text{\tiny$\leftrightarrow$}}{\mathbf{G}}_\mathrm{mm}(\mathbf{r},\omega,\mathbf{r}')\cdot\hat{\mathbf{J}}_\mathrm{m}^+(\mathbf{r}',\omega)d^3r',
 \label{eq:hfield2}
\end{align}
where we have first substituted the expression for $\hat{\mathbf{E}}^+(\mathbf{r},\omega)$
from Eq.~\eqref{eq:efield2} and
incorporated the separate $\hat{\mathbf{J}}_\mathrm{m}^+(\mathbf{r},\omega)$ term into
the integral using a suitable $\delta$-function presentation and then
defined the exchange Green's function $\overset{\text{\tiny$\leftrightarrow$}}{\mathbf{G}}_\mathrm{me}(\mathbf{r},\omega,\mathbf{r}')$
and the magnetic Green's function $\overset{\text{\tiny$\leftrightarrow$}}{\mathbf{G}}_\mathrm{mm}(\mathbf{r},\omega,\mathbf{r}')$ as
\begin{equation}
\overset{\text{\tiny$\leftrightarrow$}}{\mathbf{G}}_\mathrm{me}(\mathbf{r},\omega,\mathbf{r}')
 =\frac{\nabla_\mathbf{r}\times\overset{\text{\tiny$\leftrightarrow$}}{\mathbf{G}}_\mathrm{ee}(\mathbf{r},\omega,\mathbf{r}')}{k_0\mu(\mathbf{r},\omega)},
 \label{eq:greenme}
\end{equation}
\begin{equation}
\overset{\text{\tiny$\leftrightarrow$}}{\mathbf{G}}_\mathrm{mm}(\mathbf{r},\omega,\mathbf{r}')
 =-\frac{\nabla_\mathbf{r}\times\overset{\text{\tiny$\leftrightarrow$}}{\mathbf{G}}_\mathrm{em}(\mathbf{r},\omega,\mathbf{r}')}{k_0\mu(\mathbf{r},\omega)}
 -\overset{\text{\tiny$\leftrightarrow$}}{\mathbf{I}}\frac{\delta(\mathbf{r}-\mathbf{r}')}{k_0^2\mu(\mathbf{r},\omega)}.
 \label{eq:greenmm}
\end{equation}
By using Eqs.~\eqref{eq:greenem} and \eqref{eq:greenmm}, one also obtains an
expression of the magnetic Green's function $\overset{\text{\tiny$\leftrightarrow$}}{\mathbf{G}}_\mathrm{mm}(\mathbf{r},\omega,\mathbf{r}')$ directly in terms of the
electric Green's function $\overset{\text{\tiny$\leftrightarrow$}}{\mathbf{G}}_\mathrm{ee}(\mathbf{r},\omega,\mathbf{r}')$ as
\begin{equation}
 \overset{\text{\tiny$\leftrightarrow$}}{\mathbf{G}}_\mathrm{mm}(\mathbf{r},\omega,\mathbf{r}')
 =-\frac{\nabla_\mathbf{r}\times[\overset{\text{\tiny$\leftrightarrow$}}{\mathbf{G}}_\mathrm{ee}(\mathbf{r},\omega,\mathbf{r}')
 \times\nabla_{\mathbf{r}'}]}{k_0^2\mu(\mathbf{r},\omega)\mu(\mathbf{r}',\omega)}
 -\overset{\text{\tiny$\leftrightarrow$}}{\mathbf{I}}\frac{\delta (\mathbf{r}-\mathbf{r}')}{k_0^2\mu(\mathbf{r},\omega)}.
 \label{eq:greenmm2}
\end{equation}

\section{\label{apx:greenstratified}Green's functions for stratified media}

To gain more insight and analytical formulas directly applicable to common planar geometries,
and to partly lift the divergences associated with absorbing media,
we apply the formalism developed above for stratified media.
In the case of stratified media, it is convenient to use the plane wave representation for
the dyadic Green's functions: A point in space is denoted in the Cartesian basis
$(\hat{\mathbf{x}},\hat{\mathbf{y}},\hat{\mathbf{z}})$ by
$\mathbf{r}=x\hat{\mathbf{x}}+y\hat{\mathbf{y}}+z\hat{\mathbf{z}}=\mathbf{R}+z\hat{\mathbf{z}}$, where
$\mathbf{R}=x\hat{\mathbf{x}}+y\hat{\mathbf{y}}$
is the in-plane coordinate and the surface normals are along the $z$ coordinate.
Similarly, a wave vector of a plane wave
is denoted by $\mathbf{k}=\mathbf{K}+k_z\,\mathrm{sgn}(z-z')\hat{\mathbf{z}}$ where
the component $\mathbf{K}$ is in the $x$-$y$ plane and $k_z$ is given by
$k_z=\sqrt{k_0^2n^2-K^2}$, with $\mathrm{Im}(k_z)\ge 0$.
For convenience, we also define the unit vector $\hat{\mathbf{K}}=\mathbf{K}/K$.

The above notation is convenient since at the $x$-$y$ plane, the dyadic Green's functions of
stratified media depend only on the relative in-plane coordinate $\mathbf{R}-\mathbf{R}'$.
Therefore, in the plane-wave representation, the dyadic Green's functions
$\overset{\text{\tiny$\leftrightarrow$}}{\mathbf{G}}_{jk}(\mathbf{r},\omega,\mathbf{r}')$,
$j,k\in\{\mathrm{e,m}\}$, can be written as \cite{Tomas1995,Intravaia2015}
\begin{align}
 &\overset{\text{\tiny$\leftrightarrow$}}{\mathbf{G}}_{jk}(\mathbf{r},\omega,\mathbf{r}')\nonumber\\
 &=\frac{1}{4\pi^2}\int
 \overset{\text{\tiny$\leftrightarrow$}}{\mathbf{R}}^\mathrm{T}
 \overset{\text{\tiny$\leftrightarrow$}}{\mathbf{g}}_{jk}(z,K,\omega,z')
 \overset{\text{\tiny$\leftrightarrow$}}{\mathbf{R}}
 e^{i\mathbf{K}\cdot(\mathbf{R}-\mathbf{R}')}d^2K,\label{eq:greenfourier}
\end{align}
where the terms $\overset{\text{\tiny$\leftrightarrow$}}{\mathbf{R}}^\mathrm{T}
\overset{\text{\tiny$\leftrightarrow$}}{\mathbf{g}}_{jk}(z,K,\omega,z')
\overset{\text{\tiny$\leftrightarrow$}}{\mathbf{R}}$ are the Fourier transforms of
$\overset{\text{\tiny$\leftrightarrow$}}{\mathbf{G}}_{jk}(\mathbf{r},\omega,\mathbf{r}')$
that have been obtained by rotating the dyadic plane-wave Greens functions
$\overset{\text{\tiny$\leftrightarrow$}}{\mathbf{g}}_{jk}(z,K,\omega,z')$
calculated using the standard techniques to evaluate the fields in layered structures
as presented below. More specifically,
$\overset{\text{\tiny$\leftrightarrow$}}{\mathbf{g}}_{jk}(z,K,\omega,z')$
have been evaluated in a coordinate system where the in-plane noise components
are taken to be perpendicular and parallel to $\mathbf{K}$, and the rotation
with the rotation matrix $\overset{\text{\tiny$\leftrightarrow$}}{\mathbf{R}}$ is used
to return this convention of direction back to the coordinate system where the
direction of the dipoles does not depend on $\mathbf{K}$.

Due to the symmetry properties of the Green's functions
$\overset{\text{\tiny$\leftrightarrow$}}{\mathbf{G}}_{jk}(\mathbf{r},\omega,\mathbf{r}')$,
also $\overset{\text{\tiny$\leftrightarrow$}}{\mathbf{g}}_{jk}(z,K,\omega,z')$ obey symmetry relations.
When the field and source positions $z$ and $z'$ are interchanged,
the values of the Green's functions are changed according to the reciprocity relations as follows.
The spectral dyadic Green's functions $\overset{\text{\tiny$\leftrightarrow$}}{\mathbf{g}}_{jk}(z,K,\omega,z')$
obey the reciprocity relation
$\overset{\text{\tiny$\leftrightarrow$}}{\mathbf{g}}_{kj}(z',K,\omega,z)
=\mathrm{diag}(-1,1,-1)\overset{\text{\tiny$\leftrightarrow$}}{\mathbf{g}}_{jk}(z,K,\omega,z')^T\mathrm{diag}(-1,1,-1)$,
where $\mathrm{diag}(-1,1,-1)$ is a diagonal matrix
with diagonal elements $-1$, $1$, and $-1$.
In addition, they obey the complex conjugation relation
$\overset{\text{\tiny$\leftrightarrow$}}{\mathbf{g}}_{jk}^*(z,K,\omega,z')
=\overset{\text{\tiny$\leftrightarrow$}}{\mathbf{g}}_{jk}(z,K,-\omega,z')$.

\subsection{Multi-interface reflection and transmission coefficients}
In order to write the spectral dyadic Green's functions for a multi-interface geometry in a compact
form, we first define the multi-interface reflection and transmission coefficients
that take into account all the reflections in the geometry. First, the single-interface
reflection and transmission coefficients following from the boundary conditions
of the tangential and normal polarizations ($\sigma\in\{\parallel,\perp\}$) requiring the tangential
components of the electric and magnetic fields to be continuous
at interfaces are given for the electric and magnetic fields by
\begin{align}
 r_{e,\parallel} & =\frac{\mu_2k_{z,1}-\mu_1k_{z,2}}{\mu_2k_{z,1}+\mu_1k_{z,2}},
 \hspace{1cm} t_{e,\parallel}=\frac{2\mu_2k_{z,1}}{\mu_2k_{z,1}+\mu_1k_{z,2}},\nonumber\\
 r_{e,\perp} & =\frac{\varepsilon_1k_{z,2}-\varepsilon_2k_{z,1}}{\varepsilon_2k_{z,1}+\varepsilon_1k_{z,2}},
 \hspace{1.1cm} t_{e,\perp}=\frac{2\sqrt{\varepsilon_1/\mu_1}\,n_2k_{z,1}}{\varepsilon_2k_{z,1}+\varepsilon_1k_{z,2}},\nonumber\\
 r_{m,\parallel} & =\frac{\varepsilon_2k_{z,1}-\varepsilon_1k_{z,2}}{\varepsilon_2k_{z,1}+\varepsilon_1k_{z,2}},
 \hspace{1.1cm} t_{m,\parallel}=\frac{2\varepsilon_2k_{z,1}}{\varepsilon_2k_{z,1}+\varepsilon_1k_{z,2}},\nonumber\\
 r_{m,\perp} & =\frac{\mu_1k_{z,2}-\mu_2k_{z,1}}{\mu_2k_{z,1}+\mu_1k_{z,2}},
 \hspace{1cm} t_{m,\perp}=\frac{2\sqrt{\mu_1/\varepsilon_1}\,n_2k_{z,1}}{\mu_2k_{z,1}+\mu_1k_{z,2}},
 \label{eq:fresnel}
\end{align}
where $\varepsilon_l$, $\mu_l$, $n_l$, $k_{z,l}$, $l=1,2$, are the relative permittivities,
permeabilities, refractive indices, and the $z$ components of the wave vectors in the two materials.
The single-interface coefficients in Eq.~\eqref{eq:fresnel}
equal the conventional reflection and transmission
coefficients used, e.g., in Ref.~\cite{Paulus2000}.
In the following, with primed reflection and transmission coefficients
we denote the reflection and transmission coefficients for the incidence from
medium 2 to medium 1, and they are obtained by switching indices 1 and 2
in Eq.~\eqref{eq:fresnel}.

The multi-interface geometry is defined by interface positions $z_l$, $l=1,2,...,N$,
separating material layers with relative permittivities and permeabilities $\varepsilon_l$ and
$\mu_l$, $l=1,2,...,N+1$. The layer thicknesses are denoted by $d_l=z_l-z_{l-1}$, where $l=2,...,N$.
The multi-interface reflection and transmission coefficients $\mathcal{R}_{l,j}$
and $\mathcal{T}_{l,j}$,
which account for the multiple reflections in different medium layers,
are recursively given in terms of the single-interface reflection
and transmission coefficients as
\begin{align}
 \mathcal{R}_{l,j,\sigma} & =\frac{r_{l,j,\sigma}+\mathcal{R}_{l+1,j,\sigma}e^{2ik_{z,l+1}d_{l+1}}}{1+r_{l,j,\sigma}\mathcal{R}_{l+1,j,\sigma}e^{2ik_{z,l+1}d_{l+1}}}\label{eq:Rl},\\
 \mathcal{T}_{l,j,\sigma} & =\frac{t_{l,j,\sigma}\nu_{l+1,j,\sigma}}{\nu_{l,j,\sigma}(1-\mathcal{R}_{l-1,j,\sigma}'r_{l,j,\sigma} e^{2ik_{z,l}d_l})}\label{eq:Tl},
\end{align}
where $l=1,2,...,N$, $j\in\{\mathrm{e},\mathrm{m}\}$, $\sigma\in\{\parallel,\perp\}$
$\nu_{l,j,\sigma}=1/(1-\mathcal{R}_{l-1,j,\sigma}'\mathcal{R}_{l,j,\sigma} e^{2ik_{z,l}d_l})$,
and $\mathcal{R}_{0,j,\sigma}'=\mathcal{R}_{N+1,j,\sigma}=0$. As in the case of single-interface coefficients
in Eq.~\eqref{eq:fresnel} the primed coefficients denote the coefficients for right incidence.
The layers are indexed such that $\mathcal{R}_{l,j,\sigma}'$
corresponds to the same interface as $\mathcal{R}_{l,j,\sigma}$. The propagation coefficient
for a certain material layer of thickness $d_l$ is given as $\mathcal{P}_l=e^{ik_{z,l}d_l}$
when the transmission coefficient from layer $l'$ to layer $l>l'+1$ is recursively given by
$\mathcal{T}_{l',l,j,\sigma}=\mathcal{T}_{l',l-1,j,\sigma}\mathcal{T}_{l-1,j,\sigma}e^{ik_{z,l-1}d_{l-1}}$,
with $\mathcal{T}_{l',l'+1,j,\sigma}=\mathcal{T}_{l',j,\sigma}$, and that
from layer $l'$ to layer $l<l'-1$ by
$\mathcal{T}_{l',l,j,\sigma}'=\mathcal{T}_{l',l+1,j,\sigma}'\mathcal{T}_{l,j,\sigma}'e^{ik_{z,l+1}d_{l+1}}$,
with $\mathcal{T}_{l',l'-1,j,\sigma}'=\mathcal{T}_{l'-1,j,\sigma}'$.

\subsection{Spectral dyadic Green's functions}

Here we give a compact componentwise representation of the spectral dyadic Green's
functions for general stratified media. The presentation adapts the dyadic Green's functions
given, e.g., in Ref.~\citenum{Paulus2000} or, in the case of purely dielectric structures,
in Appendix A of Ref.~\citenum{Joulain2005}. However, the chosen presentation has a few
differences: (1) We use the orthonormal basis $(\hat{\mathbf{K}}\times\hat{\mathbf{z}},\hat{\mathbf{K}},\hat{\mathbf{z}})$,
where the in-plane noise components are taken to be perpendicular and parallel to $\mathbf{K}$.
Then, the rotation matrix $\overset{\text{\tiny$\leftrightarrow$}}{\mathbf{R}}$ in Eq.~\eqref{eq:greenfourier}
is used to return this convention of direction back to the coordinate system where the
direction of the dipoles does not depend on $\mathbf{K}$. (2) 
Instead of using the orthonormal system of complex-valued unit dyads of
Refs.~\citenum{Paulus2000} and \citenum{Joulain2005}, we write the dyadic Green's
functions as matrices. (3) We use the scaled forms of the dyadic Green's functions;
for example, $\overset{\text{\tiny$\leftrightarrow$}}{\mathbf{G}}_\mathrm{ee}(\mathbf{r},\omega,\mathbf{r}')$
is obtained as a solution to the differential equation in Eq.~\eqref{eq:greenee} instead
of the corresponding equation in Ref.~\citenum{Paulus2000}, whose right-hand side
contains an additional factor $1/\mu(\mathbf{r},\omega)$. Thus, our notation
corresponds to the notation used in the case of normal incidence in Ref.~\citenum{Partanen2016b}.

The spectral dyadic Green's functions $\overset{\text{\tiny$\leftrightarrow$}}{\mathbf{g}}_\mathrm{ee}(z,K,\omega,z')$
and $\overset{\text{\tiny$\leftrightarrow$}}{\mathbf{g}}_\mathrm{mm}(z,K,\omega,z')$, in our notation, are given in terms of the scaled
dyadic Green's functions $\overset{\text{\tiny$\leftrightarrow$}}{\boldsymbol{\xi}}_{\mathrm{e}}(z,K,\omega,z')$ and
$\overset{\text{\tiny$\leftrightarrow$}}{\boldsymbol{\xi}}_{\mathrm{m}}(z,K,\omega,z')$ as
\begin{align}
 \overset{\text{\tiny$\leftrightarrow$}}{\mathbf{g}}_\mathrm{ee}(z,K,\omega,z') &=
 \mu(z',\omega)\overset{\text{\tiny$\leftrightarrow$}}{\boldsymbol{\xi}}_{\mathrm{e}}(z,K,\omega,z')
 -\frac{\delta(z-z')}{k_0^2\varepsilon(z,\omega)}\hat{\mathbf{z}}\hat{\mathbf{z}},
 \label{eq:greendefe1}\\
 \overset{\text{\tiny$\leftrightarrow$}}{\mathbf{g}}_\mathrm{mm}(z,K,\omega,z') &=
 \varepsilon(z',\omega)\overset{\text{\tiny$\leftrightarrow$}}{\boldsymbol{\xi}}_{\mathrm{m}}(z,K,\omega,z')
 -\frac{\delta(z-z')}{k_0^2\mu(z,\omega)}\hat{\mathbf{z}}\hat{\mathbf{z}}.
 \label{eq:greendefm1}
\end{align}
The scaled dyadic Green's functions
$\overset{\text{\tiny$\leftrightarrow$}}{\boldsymbol{\xi}}_\mathrm{e}(z,\omega,z')$ and
$\overset{\text{\tiny$\leftrightarrow$}}{\boldsymbol{\xi}}_\mathrm{m}(z,\omega,z')$ in
Eqs.~\eqref{eq:greendefe1} and \eqref{eq:greendefm1} are given in
the orthonormal basis $(\hat{\mathbf{K}}\times\hat{\mathbf{z}},\hat{\mathbf{K}},\hat{\mathbf{z}})$ by
\begin{align}
 &\overset{\text{\tiny$\leftrightarrow$}}{\boldsymbol{\xi}}_{j}(z,K,\omega,z')\nonumber\\
 &=\!\!\footnotesize\left(\begin{array}{ccc}
 \!\!\!\xi_{j,\parallel}^{+}(z,\omega,z')\!\! & 0 & 0\\
 0 & \frac{k_zk_z'}{kk'}\xi_{j,\perp}^{+}(z,\omega,z') & \!i\frac{k_z K}{kk'}\frac{\partial}{k_z\partial z}\xi_{j,\perp}^{-}(z,\omega,z')\!\!\!\!\\
 0 & i\frac{K k_z'}{kk'}\frac{\partial}{k_z\partial z}\xi_{j,\perp}^{+}(z,\omega,z') & \frac{K^2}{kk'}\xi_{j,\perp}^{-}(z,\omega,z')
 \end{array}\right)\!\!,
\end{align}
where the primed and unprimed quantities correspond to the quantities
at positions $z'$ and $z$, respectively, and $\xi_{j,\sigma}^\pm(z,\omega,z')$ are the scaled scalar Green's functions,
which are presented in the case of normal incidence in Ref.~\citenum{Partanen2016b}.
For non-normal incidence, the generalization of the scaled scalar Green's functions of Ref.~\citenum{Partanen2016b}
is obtained by substituting the wave number $k$ with its $z$ component $k_z$ and the reflection
and transmission coefficients of normal incidence with the corresponding quantities
for non-normal incidence, given in Eqs.~\eqref{eq:Rl} and \eqref{eq:Tl}.
Assuming that the source point $z'$ is located in layer $l'$ ($z_{l'-1}<z'<z_{l'}$)
and the field point $z$ is located in layer $l$ ($z_{l-1}<z<z_{l}$), in
the three cases $l=l'$, $l>l'$, and $l<l'$, the scaled scalar Green's functions
are compactly given by
\begin{align}
 &\xi_{l=l',j,\sigma}^\pm(z,\omega,z')\nonumber\\
 &=\frac{i}{2k_{z,l'}}\Big(e^{ik_{z,l'}|z-z'|} \pm \nu_{l',j}\mathcal{R}_{l',j,\sigma}
 [e^{-ik_{z,l'}(z+z'-2z_{l'})}\nonumber\\
 &\hspace{0.5cm}\pm \mathcal{R}_{l'-1,j,\sigma}'e^{-ik_{z,l'}(z-z'-2d_{l'})}]\pm \nu_{l',j,\sigma}\mathcal{R}_{l'-1,j,\sigma}'\nonumber\\
 &\hspace{0.5cm}\times[e^{ik_{z,l'}(z+z'-2z_{l'-1})} \pm \mathcal{R}_{l',j,\sigma}e^{ik_{z,l'}(z-z'+2d_{l'})}]\Big),
 \label{eq:greenscalar1}
\end{align}
\begin{align}
 &\xi_{l>l',j,\sigma}^\pm(z,\omega,z')\nonumber\\
 &=\frac{i}{2k_{z,l'}}\mathcal{T}_{l',l,j,\sigma}\Big(e^{ik_{z,l'}(z_{l'}-z')} \pm \nu_{l',j,\sigma}\mathcal{R}_{l'-1,j,\sigma}'\nonumber\\
 &\hspace{0.5cm}\times[e^{ik_{z,l'}(z'+d_{l'}-z_{l'-1})}\pm \mathcal{R}_{l',j,\sigma}e^{ik_{z,l'}(2d_{l'}-z'+z_{l'})}]\Big)\nonumber\\
 &\hspace{0.5cm}\times\Big(e^{ik_{z,l}(z-z_{l-1})} \pm \mathcal{R}_{l,j,\sigma}e^{-ik_{z,l}(z-z_{l-1}-2d_{l})}\Big),
 \label{eq:greenscalar2}
\end{align}
\begin{align}
 &\xi_{l<l',j,\sigma}^\pm(z,\omega,z')\nonumber\\
 &=\frac{i}{2k_{z,l'}}\mathcal{T}_{l',l,j,\sigma}'\Big(e^{ik_{z,l'}(z'-z_{l'-1})} \pm \nu_{l',j,\sigma}\mathcal{R}_{l',j,\sigma}\nonumber\\
 &\hspace{0.4cm}\times[e^{-ik_{z,l'}(z'-2d_{l'}-z_{l'-1})}\!\pm\!\mathcal{R}_{l'-1,j,\sigma}'e^{ik_{z,l'}(z'+2d_{l'}-z_{l'-1})}]\Big)\nonumber\\
 &\hspace{0.4cm}\times\Big(e^{-ik_{z,l}(z-z_{l})} \pm \mathcal{R}_{l-1,j,\sigma}'e^{ik_{z,l}(z+d_{l}-z_{l-1})}\Big).
 \label{eq:greenscalar3}
\end{align}

The spectral dyadic Green's functions $\overset{\text{\tiny$\leftrightarrow$}}{\mathbf{g}}_\mathrm{me}(z,K,\omega,z')$
and $\overset{\text{\tiny$\leftrightarrow$}}{\mathbf{g}}_\mathrm{em}(z,K,\omega,z')$ following
from Eqs.~\eqref{eq:greenem}, \eqref{eq:greenme}, and \eqref{eq:greenfourier} are, respectively, presented in terms of the scaled
dyadic exchange Green's functions $\overset{\text{\tiny$\leftrightarrow$}}{\boldsymbol{\xi}}_{\mathrm{ex,e}}(z,K,\omega,z')$
and $\overset{\text{\tiny$\leftrightarrow$}}{\boldsymbol{\xi}}_{\mathrm{ex,m}}(z,K,\omega,z')$ as
\begin{align}
 \overset{\text{\tiny$\leftrightarrow$}}{\mathbf{g}}_\mathrm{me}(z,K,\omega,z') &=\frac{\mu(z',\omega)}{\mu(z,\omega)}
\overset{\text{\tiny$\leftrightarrow$}}{\boldsymbol{\xi}}_{\mathrm{ex,e}}(z,K,\omega,z'),
 \label{eq:greendefe2}\\
 \overset{\text{\tiny$\leftrightarrow$}}{\mathbf{g}}_\mathrm{em}(z,K,\omega,z') &=-\frac{\varepsilon(z',\omega)}{\varepsilon(z,\omega)}
 \overset{\text{\tiny$\leftrightarrow$}}{\boldsymbol{\xi}}_{\mathrm{ex,m}}(z,K,\omega,z').
 \label{eq:greendefm2}
\end{align}
The scaled dyadic exchange Green's functions
$\overset{\text{\tiny$\leftrightarrow$}}{\boldsymbol{\xi}}_{\mathrm{ex,e}}(z,K,\omega,z')$
and $\overset{\text{\tiny$\leftrightarrow$}}{\boldsymbol{\xi}}_{\mathrm{ex,m}}(z,K,\omega,z')$
are given in terms of the scaled scalar Green's functions in
Eqs.~\eqref{eq:greenscalar1}--\eqref{eq:greenscalar3} by
\begin{align}
 &\overset{\text{\tiny$\leftrightarrow$}}{\boldsymbol{\xi}}_{\mathrm{ex},j}(z,K,\omega,z')\nonumber\\
 &\!=\!\!\footnotesize\left(\begin{array}{ccc}
 0 & \!\!\!\!-\frac{k_z'k}{k_z k'}\frac{\partial}{k_0\partial z}\xi_{j,\perp}^{+}(z,\omega,z') & \!i\frac{Kk}{k_0k'}\xi_{j,\perp}^{-}(z,\omega,z')\!\!\\
 \!\!\!\frac{\partial}{k_0\partial z}\xi_{j,\parallel}^{+}(z,\omega,z') & 0 & 0\\
 \!\!\!-i\frac{K}{k_0}\xi_{j,\parallel}^{+}(z,\omega,z') & 0 & 0
 \end{array}\right)\!\!.
\end{align}

\section{\label{apx:densities}Derivation of the densities of states}

\subsection{\label{apx:nldos}Nonlocal densities of states}

The time-domain field operators are obtained from the frequency-domain operators by
Fourier transforms. For example, the time-domain electric field operator is given by
\begin{equation}
 \hat{\mathbf{E}}(\mathbf{r},t)\!=\!\frac{1}{2\pi}\!\int_0^\infty\!\!\hat{\mathbf{E}}^+(\mathbf{r},\omega)e^{-i\omega t}d\omega
 +\frac{1}{2\pi}\!\int_0^\infty\!\!\hat{\mathbf{E}}^-(\mathbf{r},\omega)e^{i\omega t}d\omega,
\end{equation}
where $\hat{\mathbf{E}}^-(\mathbf{r},\omega)$ is the negative-frequency part
obtained by a Hermitian conjugate of the positive-frequency part $\hat{\mathbf{E}}^+(\mathbf{r},\omega)$
in Eq.~\eqref{eq:efield}.

The frequency-space correlation functions are given by
\begin{align}
 &\langle\hat{\mathbf{E}}^{-}(\mathbf{r},\omega)\cdot\hat{\mathbf{E}}^+(\mathbf{r},\omega')\rangle\nonumber\\
  &=\mu_0^2\omega\omega'\int\langle\hat{\mathbf{J}}_\mathrm{e}^\dag(\mathbf{r}',\omega)\cdot\overset{\text{\tiny$\leftrightarrow$}}{\mathbf{G}}_\mathrm{ee}^\dag(\mathbf{r},\omega,\mathbf{r}')\nonumber\\
  &\hspace{0.5cm}\cdot\overset{\text{\tiny$\leftrightarrow$}}{\mathbf{G}}_\mathrm{ee}(\mathbf{r},\omega',\mathbf{r}'')\cdot\hat{\mathbf{J}}_\mathrm{e}(\mathbf{r}'',\omega')\rangle d^3r'd^3r''\nonumber\\
&\hspace{0.5cm}+k_0^2\int\langle\hat{\mathbf{J}}_\mathrm{m}^\dag(\mathbf{r}',\omega)\cdot\overset{\text{\tiny$\leftrightarrow$}}{\mathbf{G}}_\mathrm{em}^\dag(\mathbf{r},\omega,\mathbf{r}')\nonumber\\
&\hspace{0.5cm}\cdot\overset{\text{\tiny$\leftrightarrow$}}{\mathbf{G}}_\mathrm{em}(\mathbf{r},\omega',\mathbf{r}'')\cdot\hat{\mathbf{J}}_\mathrm{m}(\mathbf{r}'',\omega')\rangle d^3r'd^3r''\nonumber\\
  &=\delta(\omega-\omega')\mu_0^2\omega^2\int|j_\mathrm{0,e}(\mathbf{r}',\omega)|^2\nonumber\\
&\hspace{0.5cm}\times\mathrm{Tr}[\overset{\text{\tiny$\leftrightarrow$}}{\mathbf{G}}_\mathrm{ee}^\dag(\mathbf{r},\omega,\mathbf{r}')\cdot
 \overset{\text{\tiny$\leftrightarrow$}}{\mathbf{G}}_\mathrm{ee}(\mathbf{r},\omega,\mathbf{r}')]\langle\hat\eta(\mathbf{r}',\omega)\rangle d^3r'\nonumber\\
&\hspace{0.5cm}+\delta(\omega-\omega')k_0^2\int|j_\mathrm{0,m}(\mathbf{r}',\omega)|^2\nonumber\\
&\hspace{0.5cm}\times\mathrm{Tr}[\overset{\text{\tiny$\leftrightarrow$}}{\mathbf{G}}_\mathrm{em}^\dag(\mathbf{r},\omega,\mathbf{r}')\cdot
 \overset{\text{\tiny$\leftrightarrow$}}{\mathbf{G}}_\mathrm{em}(\mathbf{r},\omega,\mathbf{r}')]\langle\hat\eta(\mathbf{r}',\omega)\rangle d^3r',
\end{align}
\begin{align}
 &\langle\hat{\mathbf{E}}^{+}(\mathbf{r},\omega)\cdot\hat{\mathbf{E}}^-(\mathbf{r},\omega')\rangle\nonumber\\
  &=\mu_0^2\omega\omega'\int\langle\hat{\mathbf{J}}_\mathrm{e}(\mathbf{r}',\omega)\cdot\overset{\text{\tiny$\leftrightarrow$}}{\mathbf{G}}_\mathrm{ee}(\mathbf{r},\omega,\mathbf{r}')\nonumber\\
  &\hspace{0.5cm}\cdot\overset{\text{\tiny$\leftrightarrow$}}{\mathbf{G}}_\mathrm{ee}^\dag(\mathbf{r},\omega',\mathbf{r}'')\cdot\hat{\mathbf{J}}_\mathrm{e}^\dag(\mathbf{r}'',\omega')\rangle d^3r'd^3r''\nonumber\\
&\hspace{0.5cm}+k_0^2\int\langle\hat{\mathbf{J}}_\mathrm{m}(\mathbf{r}',\omega)\cdot\overset{\text{\tiny$\leftrightarrow$}}{\mathbf{G}}_\mathrm{em}(\mathbf{r},\omega,\mathbf{r}')\nonumber\\
&\hspace{0.5cm}\cdot\overset{\text{\tiny$\leftrightarrow$}}{\mathbf{G}}_\mathrm{em}^\dag(\mathbf{r},\omega',\mathbf{r}'')\cdot\hat{\mathbf{J}}_\mathrm{m}^\dag(\mathbf{r}'',\omega')\rangle d^3r'd^3r''\nonumber\\
  &=\delta(\omega-\omega')\mu_0^2\omega^2\int|j_\mathrm{0,e}(\mathbf{r}',\omega)|^2\nonumber\\
&\hspace{0.5cm}\times\mathrm{Tr}[\overset{\text{\tiny$\leftrightarrow$}}{\mathbf{G}}_\mathrm{ee}(\mathbf{r},\omega,\mathbf{r}')\cdot
 \overset{\text{\tiny$\leftrightarrow$}}{\mathbf{G}}_\mathrm{ee}^\dag(\mathbf{r},\omega,\mathbf{r}')][\langle\hat\eta(\mathbf{r}',\omega)\rangle+1]d^3r'\nonumber\\
&\hspace{0.5cm}+\delta(\omega-\omega')k_0^2\int|j_\mathrm{0,m}(\mathbf{r}',\omega)|^2\nonumber\\
&\hspace{0.5cm}\times\mathrm{Tr}[\overset{\text{\tiny$\leftrightarrow$}}{\mathbf{G}}_\mathrm{em}(\mathbf{r},\omega,\mathbf{r}')\cdot
 \overset{\text{\tiny$\leftrightarrow$}}{\mathbf{G}}_\mathrm{em}^\dag(\mathbf{r},\omega,\mathbf{r}')][\langle\hat\eta(\mathbf{r}',\omega)\rangle+1]d^3r'.
\end{align}

In the time domain, we have
\begin{align}
 &\langle\hat{\mathbf{E}}(\mathbf{r},t)^2\rangle\nonumber\\
 &=\frac{1}{4\pi^2}\int_0^\infty\!\int_0^\infty\langle\hat{\mathbf{E}}^-(\mathbf{r},\omega)\cdot\hat{\mathbf{E}}^+(\mathbf{r},\omega')\rangle e^{i(\omega-\omega')t}d\omega d\omega'\nonumber\\
 &+\frac{1}{4\pi^2}\int_0^\infty\!\int_0^\infty\langle\hat{\mathbf{E}}^+(\mathbf{r},\omega)\cdot\hat{\mathbf{E}}^-(\mathbf{r},\omega')\rangle e^{i(\omega'-\omega)t}d\omega d\omega',
\end{align}
which then becomes
\begin{align}
 &\langle\hat{\mathbf{E}}(\mathbf{r},t)^2\rangle\nonumber\\
 &=\int_0^\infty\!\!\!\int\!\frac{\mu_0^2\omega^2}{2\pi^2}|j_\mathrm{0,e}(\mathbf{r}',\omega)|^2
\mathrm{Tr}[\overset{\text{\tiny$\leftrightarrow$}}{\mathbf{G}}_\mathrm{ee}(\mathbf{r},\omega,\mathbf{r}')\cdot
 \overset{\text{\tiny$\leftrightarrow$}}{\mathbf{G}}_\mathrm{ee}^\dag(\mathbf{r},\omega,\mathbf{r}')]\nonumber\\
&\hspace{0.5cm}\times\Big(\langle\hat\eta(\mathbf{r}',\omega)\rangle+\frac{1}{2}\Big)d^3r'd\omega\nonumber\\
&\hspace{0.4cm}+\!\int_0^\infty\!\!\!\int\!\!\frac{k_0^2}{2\pi^2}|j_\mathrm{0,m}(\mathbf{r}',\omega)|^2
\mathrm{Tr}[\overset{\text{\tiny$\leftrightarrow$}}{\mathbf{G}}_\mathrm{em}(\mathbf{r},\omega,\mathbf{r}')\cdot
 \overset{\text{\tiny$\leftrightarrow$}}{\mathbf{G}}_\mathrm{em}^\dag(\mathbf{r},\omega,\mathbf{r}')]\nonumber\\
&\hspace{0.5cm}\times\Big(\langle\hat\eta(\mathbf{r}',\omega)\rangle+\frac{1}{2}\Big)d^3r'd\omega.
\end{align}

Using $|j_\mathrm{0,e}(\mathbf{r}',\omega)|^2=4\pi\hbar\omega^2\varepsilon_0\varepsilon_\mathrm{i}(\mathbf{r}',\omega)$ and
$|j_\mathrm{0,m}(\mathbf{r}',\omega)|^2=4\pi\hbar\omega^2\mu_0\mu_\mathrm{i}(\mathbf{r}',\omega)$ gives
\begin{align}
 &\langle\hat{\mathbf{E}}(\mathbf{r},t)^2\rangle\nonumber\\
 &=\int_0^\infty\!\!\int\frac{2\hbar\omega^4\mu_0}{\pi c^2}\Big(\varepsilon_\mathrm{i}(\mathbf{r}',\omega)
\mathrm{Tr}[\overset{\text{\tiny$\leftrightarrow$}}{\mathbf{G}}_\mathrm{ee}(\mathbf{r},\omega,\mathbf{r}')\cdot
 \overset{\text{\tiny$\leftrightarrow$}}{\mathbf{G}}_\mathrm{ee}^\dag(\mathbf{r},\omega,\mathbf{r}')]\nonumber\\
&\hspace{0.5cm}+\mu_\mathrm{i}(\mathbf{r}',\omega)
\mathrm{Tr}[\overset{\text{\tiny$\leftrightarrow$}}{\mathbf{G}}_\mathrm{em}(\mathbf{r},\omega,\mathbf{r}')\cdot
 \overset{\text{\tiny$\leftrightarrow$}}{\mathbf{G}}_\mathrm{em}^\dag(\mathbf{r},\omega,\mathbf{r}')]\Big)\nonumber\\
&\hspace{0.5cm}\times\Big(\langle\hat\eta(\mathbf{r}',\omega)\rangle+\frac{1}{2}\Big)d^3r'd\omega.
\label{eq:apxefluct}
\end{align}
This allows defining the NLDOS for the electric field as
\begin{align}
 &\rho_\mathrm{NL,e}(\mathbf{r},\omega,\mathbf{r}')\nonumber\\
 &=\frac{2\omega^3}{\pi c^4}\Big(\varepsilon_\mathrm{i}(\mathbf{r}',\omega)
\mathrm{Tr}[\overset{\text{\tiny$\leftrightarrow$}}{\mathbf{G}}_\mathrm{ee}(\mathbf{r},\omega,\mathbf{r}')\cdot
 \overset{\text{\tiny$\leftrightarrow$}}{\mathbf{G}}_\mathrm{ee}^\dag(\mathbf{r},\omega,\mathbf{r}')]\nonumber\\
&\hspace{0.5cm}+\mu_\mathrm{i}(\mathbf{r}',\omega)
\mathrm{Tr}[\overset{\text{\tiny$\leftrightarrow$}}{\mathbf{G}}_\mathrm{em}(\mathbf{r},\omega,\mathbf{r}')\cdot
 \overset{\text{\tiny$\leftrightarrow$}}{\mathbf{G}}_\mathrm{em}^\dag(\mathbf{r},\omega,\mathbf{r}')]\Big).
\label{eq:apxenldos}
\end{align}
Corresponding equations can be written for the magnetic field.
The NLDOS of the magnetic field is then given by
\begin{align}
 &\rho_\mathrm{NL,m}(\mathbf{r},\omega,\mathbf{r}')\nonumber\\
 &=\frac{2\omega^3}{\pi c^4}\Big(\varepsilon_\mathrm{i}(\mathbf{r}',\omega)
\mathrm{Tr}[\overset{\text{\tiny$\leftrightarrow$}}{\mathbf{G}}_\mathrm{me}(\mathbf{r},\omega,\mathbf{r}')\cdot
 \overset{\text{\tiny$\leftrightarrow$}}{\mathbf{G}}_\mathrm{me}^\dag(\mathbf{r},\omega,\mathbf{r}')]\nonumber\\
&\hspace{0.5cm}+\mu_\mathrm{i}(\mathbf{r}',\omega)
\mathrm{Tr}[\overset{\text{\tiny$\leftrightarrow$}}{\mathbf{G}}_\mathrm{mm}(\mathbf{r},\omega,\mathbf{r}')\cdot
 \overset{\text{\tiny$\leftrightarrow$}}{\mathbf{G}}_\mathrm{mm}^\dag(\mathbf{r},\omega,\mathbf{r}')]\Big).
\label{eq:apxhnldos}
\end{align}

\subsection{\label{apx:ifdos}Interference density of states}

For an optical mode the
quantum optical Poynting vector is defined as a normal ordered operator
in terms of the positive- and negative-frequency parts of the electric 
and magnetic field operators as
$\hat{\mathbf{S}}(\mathbf{r},t)=:\!\hat{\mathbf{E}}(\mathbf{r},t)\times\hat{\mathbf{H}}(\mathbf{r},t)\!:=\hat{\mathbf{E}}^-(\mathbf{r},t)\times\hat{\mathbf{H}}^+(\mathbf{r},t)-\hat{\mathbf{H}}^-(\mathbf{r},t)\times\hat{\mathbf{E}}^+(\mathbf{r},t)$
\cite{Loudon2000}.
Substituting the time-space forms of the electric and magnetic field operators
in Eqs.~\eqref{eq:efield} and \eqref{eq:hfield} gives
\begin{align}
 &\langle\hat{\mathbf{S}}(\mathbf{r},t)\rangle\nonumber\\
 &=\frac{1}{4\pi^2}\int_0^\infty\int_0^\infty\langle\hat{\mathbf{E}}^-(\mathbf{r},\omega)\times\hat{\mathbf{H}}^+(\mathbf{r},\omega')\rangle e^{i(\omega-\omega')t}d\omega d\omega'\nonumber\\
 &-\frac{1}{4\pi^2}\int_0^\infty\int_0^\infty\langle\hat{\mathbf{H}}^-(\mathbf{r},\omega)\times\hat{\mathbf{E}}^+(\mathbf{r},\omega')\rangle e^{i(\omega-\omega')t}d\omega d\omega'
\end{align}

The frequency-space correlation functions are given by
\begin{align}
 &\langle\hat{\mathbf{E}}^{-}(\mathbf{r},\omega)\times\hat{\mathbf{H}}^+(\mathbf{r},\omega')\rangle\nonumber\\
  &=-i\omega\mu_0k_0\int\langle[\hat{\mathbf{J}}_\mathrm{e}^\dag(\mathbf{r}',\omega)\cdot\overset{\text{\tiny$\leftrightarrow$}}{\mathbf{G}}_\mathrm{ee}^\dag(\mathbf{r},\omega,\mathbf{r}')]\nonumber\\
  &\hspace{0.5cm}\times[\overset{\text{\tiny$\leftrightarrow$}}{\mathbf{G}}_\mathrm{me}(\mathbf{r},\omega',\mathbf{r}'')\cdot\hat{\mathbf{J}}_\mathrm{e}(\mathbf{r}'',\omega')]\rangle d^3r'd^3r''\nonumber\\
&\hspace{0.5cm}+i\omega'\varepsilon_0k_0\int\langle[\hat{\mathbf{J}}_\mathrm{m}^\dag(\mathbf{r}',\omega)\cdot\overset{\text{\tiny$\leftrightarrow$}}{\mathbf{G}}_\mathrm{em}^\dag(\mathbf{r},\omega,\mathbf{r}')]\nonumber\\
&\hspace{0.5cm}\times[\overset{\text{\tiny$\leftrightarrow$}}{\mathbf{G}}_\mathrm{mm}(\mathbf{r},\omega',\mathbf{r}'')\cdot\hat{\mathbf{J}}_\mathrm{m}(\mathbf{r}'',\omega')]\rangle d^3r'd^3r''\nonumber\\
  &=-\delta(\omega-\omega')i\omega\mu_0k_0\int|j_\mathrm{0,e}(\mathbf{r}',\omega)|^2\nonumber\\
&\hspace{0.5cm}\times\mathrm{Tr}[\overset{\text{\tiny$\leftrightarrow$}}{\mathbf{G}}_\mathrm{ee}^\dag(\mathbf{r},\omega,\mathbf{r}')\times
 \overset{\text{\tiny$\leftrightarrow$}}{\mathbf{G}}_\mathrm{me}(\mathbf{r},\omega,\mathbf{r}')]\langle\hat\eta(\mathbf{r}',\omega)\rangle d^3r'\nonumber\\
&\hspace{0.5cm}+\delta(\omega-\omega')i\omega\varepsilon_0k_0\int|j_\mathrm{0,m}(\mathbf{r}',\omega)|^2\nonumber\\
&\hspace{0.5cm}\times\mathrm{Tr}[\overset{\text{\tiny$\leftrightarrow$}}{\mathbf{G}}_\mathrm{em}^\dag(\mathbf{r},\omega,\mathbf{r}')\times
 \overset{\text{\tiny$\leftrightarrow$}}{\mathbf{G}}_\mathrm{mm}(\mathbf{r},\omega,\mathbf{r}')]\langle\hat\eta(\mathbf{r}',\omega)\rangle d^3r',
\end{align}
\begin{align}
 &\langle\hat{\mathbf{H}}^{-}(\mathbf{r},\omega)\times\hat{\mathbf{E}}^+(\mathbf{r},\omega')\rangle\nonumber\\
  &=-i\omega\varepsilon_0k_0\int\langle[\hat{\mathbf{J}}_\mathrm{m}^\dag(\mathbf{r}',\omega)\cdot\overset{\text{\tiny$\leftrightarrow$}}{\mathbf{G}}_\mathrm{mm}^\dag(\mathbf{r},\omega,\mathbf{r}')]\nonumber\\
  &\hspace{0.5cm}\times[\overset{\text{\tiny$\leftrightarrow$}}{\mathbf{G}}_\mathrm{em}(\mathbf{r},\omega',\mathbf{r}'')\cdot\hat{\mathbf{J}}_\mathrm{m}(\mathbf{r}'',\omega')]\rangle d^3r'd^3r''\nonumber\\
&\hspace{0.5cm}+i\omega'\mu_0k_0\int\langle[\hat{\mathbf{J}}_\mathrm{e}^\dag(\mathbf{r}',\omega)\cdot\overset{\text{\tiny$\leftrightarrow$}}{\mathbf{G}}_\mathrm{me}^\dag(\mathbf{r},\omega,\mathbf{r}')]\nonumber\\
&\hspace{0.5cm}\times[\overset{\text{\tiny$\leftrightarrow$}}{\mathbf{G}}_\mathrm{ee}(\mathbf{r},\omega',\mathbf{r}'')\cdot\hat{\mathbf{J}}_\mathrm{e}(\mathbf{r}'',\omega')]\rangle d^3r'd^3r''\nonumber\\
  &=-\delta(\omega-\omega')i\omega\varepsilon_0k_0\int|j_\mathrm{0,m}(\mathbf{r}',\omega)|^2\nonumber\\
&\hspace{0.5cm}\times\mathrm{Tr}[\overset{\text{\tiny$\leftrightarrow$}}{\mathbf{G}}_\mathrm{mm}^\dag(\mathbf{r},\omega,\mathbf{r}')\times
 \overset{\text{\tiny$\leftrightarrow$}}{\mathbf{G}}_\mathrm{em}(\mathbf{r},\omega,\mathbf{r}')]\langle\hat\eta(\mathbf{r}',\omega)\rangle d^3r'\nonumber\\
&\hspace{0.5cm}+\delta(\omega-\omega')i\omega\mu_0k_0\int|j_\mathrm{0,e}(\mathbf{r}',\omega)|^2\nonumber\\
&\hspace{0.5cm}\times\mathrm{Tr}[\overset{\text{\tiny$\leftrightarrow$}}{\mathbf{G}}_\mathrm{me}^\dag(\mathbf{r},\omega,\mathbf{r}')\times
 \overset{\text{\tiny$\leftrightarrow$}}{\mathbf{G}}_\mathrm{ee}(\mathbf{r},\omega,\mathbf{r}')]\langle\hat\eta(\mathbf{r}',\omega)\rangle d^3r'.
\end{align}
Here $\mathrm{Tr}[\overset{\text{\tiny$\leftrightarrow$}}{\mathbf{G}}_{jj}^\dag(\mathbf{r},\omega,\mathbf{r}')\times
 \overset{\text{\tiny$\leftrightarrow$}}{\mathbf{G}}_{kj}(\mathbf{r},\omega,\mathbf{r}')]
 =\sum_\sigma[\hat{\mathbf{e}}_\sigma\cdot\overset{\text{\tiny$\leftrightarrow$}}{\mathbf{G}}_{jj}^\dag(\mathbf{r},\omega,\mathbf{r}')]\times
 [\overset{\text{\tiny$\leftrightarrow$}}{\mathbf{G}}_{kj}(\mathbf{r},\omega,\mathbf{r}')\cdot\hat{\mathbf{e}}_\sigma]$,
which is a vector, in contrast to the conventional trace of a matrix.
The Poynting vector then becomes
\begin{align}
 &\langle\hat{\mathbf{S}}(\mathbf{r},t)\rangle\nonumber\\
 &=\frac{1}{4\pi^2}\int_0^\infty\int\Big(
 -i\omega\mu_0k_0|j_\mathrm{0,e}(\mathbf{r}',\omega)|^2\nonumber\\
 &\hspace{0.5cm}\times\mathrm{Tr}\Big[\overset{\text{\tiny$\leftrightarrow$}}{\mathbf{G}}_\mathrm{ee}^\dag(\mathbf{r},\omega,\mathbf{r}')\times
 \overset{\text{\tiny$\leftrightarrow$}}{\mathbf{G}}_\mathrm{me}(\mathbf{r},\omega,\mathbf{r}')\nonumber\\
 &\hspace{0.5cm}+\overset{\text{\tiny$\leftrightarrow$}}{\mathbf{G}}_\mathrm{me}^\dag(\mathbf{r},\omega,\mathbf{r}')\times
 \overset{\text{\tiny$\leftrightarrow$}}{\mathbf{G}}_\mathrm{ee}(\mathbf{r},\omega,\mathbf{r}')\Big]\nonumber\\
 &\hspace{0.5cm}+i\omega\varepsilon_0k_0|j_\mathrm{0,m}(\mathbf{r}',\omega)|^2\nonumber\\
&\hspace{0.5cm}\times\mathrm{Tr}\Big[\overset{\text{\tiny$\leftrightarrow$}}{\mathbf{G}}_\mathrm{em}^\dag(\mathbf{r},\omega,\mathbf{r}')\times
 \overset{\text{\tiny$\leftrightarrow$}}{\mathbf{G}}_\mathrm{mm}(\mathbf{r},\omega,\mathbf{r}')\nonumber\\
 &\hspace{0.5cm}+\overset{\text{\tiny$\leftrightarrow$}}{\mathbf{G}}_\mathrm{mm}^\dag(\mathbf{r},\omega,\mathbf{r}')\times
 \overset{\text{\tiny$\leftrightarrow$}}{\mathbf{G}}_\mathrm{em}(\mathbf{r},\omega,\mathbf{r}')\Big]
 \Big)\langle\hat\eta(\mathbf{r}',\omega)\rangle d^3r'd\omega\nonumber\\
 &=\frac{1}{2\pi^2}\int_0^\infty\int\Big(
 -\omega\mu_0k_0|j_\mathrm{0,e}(\mathbf{r}',\omega)|^2\nonumber\\
 &\hspace{0.5cm}\times\mathrm{Im}\Big[\mathrm{Tr}[\overset{\text{\tiny$\leftrightarrow$}}{\mathbf{G}}_\mathrm{ee}(\mathbf{r},\omega,\mathbf{r}')\times
 \overset{\text{\tiny$\leftrightarrow$}}{\mathbf{G}}_\mathrm{me}^\dag(\mathbf{r},\omega,\mathbf{r}')]\Big]\nonumber\\
 &\hspace{0.5cm}+\omega\varepsilon_0k_0|j_\mathrm{0,m}(\mathbf{r}',\omega)|^2\nonumber\\
 &\hspace{0.5cm}\times\mathrm{Im}\Big[\mathrm{Tr}[\overset{\text{\tiny$\leftrightarrow$}}{\mathbf{G}}_\mathrm{mm}(\mathbf{r},\omega,\mathbf{r}')\times
 \overset{\text{\tiny$\leftrightarrow$}}{\mathbf{G}}_\mathrm{em}^\dag(\mathbf{r},\omega,\mathbf{r}')]\Big]\Big)\nonumber\\
 &\hspace{0.5cm}\times\langle\hat\eta(\mathbf{r}',\omega)\rangle d^3r'd\omega.
\end{align}

Using $|j_\mathrm{0,e}(\mathbf{r}',\omega)|^2=4\pi\hbar\omega^2\varepsilon_0\varepsilon_\mathrm{i}(\mathbf{r}',\omega)$ and
$|j_\mathrm{0,m}(\mathbf{r}',\omega)|^2=4\pi\hbar\omega^2\mu_0\mu_\mathrm{i}(\mathbf{r}',\omega)$ gives
\begin{align}
 &\langle\hat{\mathbf{S}}(\mathbf{r},t)\rangle\nonumber\\
 &=\int_0^\infty\int\frac{2\hbar\omega^4}{\pi c^3}\nonumber\\
 &\hspace{0.5cm}\times\Big(\mu_\mathrm{i}(\mathbf{r}',\omega)
 \mathrm{Im}\Big[\mathrm{Tr}[\overset{\text{\tiny$\leftrightarrow$}}{\mathbf{G}}_\mathrm{mm}(\mathbf{r},\omega,\mathbf{r}')\times
 \overset{\text{\tiny$\leftrightarrow$}}{\mathbf{G}}_\mathrm{em}^\dag(\mathbf{r},\omega,\mathbf{r}')]\Big]\nonumber\\
 &\hspace{0.5cm}-\varepsilon_\mathrm{i}(\mathbf{r}',\omega)
 \mathrm{Im}\Big[\mathrm{Tr}[\overset{\text{\tiny$\leftrightarrow$}}{\mathbf{G}}_\mathrm{ee}(\mathbf{r},\omega,\mathbf{r}')\times
 \overset{\text{\tiny$\leftrightarrow$}}{\mathbf{G}}_\mathrm{me}^\dag(\mathbf{r},\omega,\mathbf{r}')]\Big]\Big)\nonumber\\
 &\hspace{0.5cm}\times\langle\hat\eta(\mathbf{r}',\omega)\rangle d^3r'd\omega.
 \label{eq:apxpoynting}
\end{align}
This allows defining the IFDOS as
\begin{align}
 &\boldsymbol{\rho}_\mathrm{IF}(\mathbf{r},\omega,\mathbf{r}')\nonumber\\
 &=\frac{2\omega^3n_\mathrm{r}(\mathbf{r},\omega)}{\pi c^4}\nonumber\\
 &\hspace{0.5cm}\times\Big(\mu_\mathrm{i}(\mathbf{r}',\omega)
 \mathrm{Im}\Big[\mathrm{Tr}[\overset{\text{\tiny$\leftrightarrow$}}{\mathbf{G}}_\mathrm{mm}(\mathbf{r},\omega,\mathbf{r}')\times
 \overset{\text{\tiny$\leftrightarrow$}}{\mathbf{G}}_\mathrm{em}^\dag(\mathbf{r},\omega,\mathbf{r}')]\Big]\nonumber\\
 &\hspace{0.5cm}-\varepsilon_\mathrm{i}(\mathbf{r}',\omega)
 \mathrm{Im}\Big[\mathrm{Tr}[\overset{\text{\tiny$\leftrightarrow$}}{\mathbf{G}}_\mathrm{ee}(\mathbf{r},\omega,\mathbf{r}')\times
 \overset{\text{\tiny$\leftrightarrow$}}{\mathbf{G}}_\mathrm{me}^\dag(\mathbf{r},\omega,\mathbf{r}')]\Big]\Big),
\end{align}
where $n_\mathrm{r}(\mathbf{r},\omega)$ is the real part of the refractive index.

\section{\label{apx:stratifieddos}Densities of states for stratified media}

Here we present the densities of states for stratified media by using 
the components $g_{jk}^{\alpha\beta}$,
$\alpha,\beta\in\{1,2,3\}$, of the matrix representations of the spectral dyadic Green's functions
$\overset{\text{\tiny$\leftrightarrow$}}{\mathbf{g}}_{jk}$.

\subsection{Nonlocal densities of states}

Using Eq.~\eqref{eq:apxefluct} with $\varepsilon(\mathbf{r}',\omega)=\varepsilon(z',\omega)$,
$\mu(\mathbf{r}',\omega)=\mu(z',\omega)$,
$\langle\hat\eta(\mathbf{r}',\omega)\rangle=\langle\hat\eta(z',\omega)\rangle$, and
\begin{align}
 &\int\mathrm{Tr}[\overset{\text{\tiny$\leftrightarrow$}}{\mathbf{G}}_{jk}^\dag(\mathbf{r},\omega,\mathbf{r}')\cdot
 \overset{\text{\tiny$\leftrightarrow$}}{\mathbf{G}}_{jk}(\mathbf{r},\omega,\mathbf{r}')]d^2R'\nonumber\\
 &=\frac{1}{4\pi^2}
 \int\mathrm{Tr}[\overset{\text{\tiny$\leftrightarrow$}}{\mathbf{g}}_{jk}^\dag(z,K,\omega,z')\cdot
 \overset{\text{\tiny$\leftrightarrow$}}{\mathbf{g}}_{jk}(z,K,\omega,z')]d^2K,
\end{align}
where $j,k\in\{\mathrm{e,m}\}$, gives
\begin{align}
 &\langle\hat{\mathbf{E}}(\mathbf{r},t)^2\rangle\nonumber\\
 &=\int\int_0^\infty\int_{-\infty}^\infty\frac{\hbar\omega^4\mu_0}{2\pi^3c^2}\nonumber\\
 &\hspace{0.5cm}\times\Big(\varepsilon_\mathrm{i}(z',\omega)
 \mathrm{Tr}[\overset{\text{\tiny$\leftrightarrow$}}{\mathbf{g}}_{\mathrm{ee}}^\dag(z,K,\omega,z')\cdot
 \overset{\text{\tiny$\leftrightarrow$}}{\mathbf{g}}_{\mathrm{ee}}(z,K,\omega,z')]\nonumber\\
&\hspace{0.5cm}+\mu_\mathrm{i}(z',\omega)
 \mathrm{Tr}[\overset{\text{\tiny$\leftrightarrow$}}{\mathbf{g}}_{\mathrm{em}}^\dag(z,K,\omega,z')\cdot
 \overset{\text{\tiny$\leftrightarrow$}}{\mathbf{g}}_{\mathrm{em}}(z,K,\omega,z')]\Big)\nonumber\\
 &\hspace{0.5cm}\times\Big(\langle\hat\eta(z',\omega)\rangle+\frac{1}{2}\Big)dz'd\omega d^2K.
\end{align}
Then, the NLDOS for the electric field can be written as
\begin{align}
 &\rho_\mathrm{NL,e}(z,K,\omega,z')\nonumber\\
 &=\frac{\omega^3}{2\pi^3c^4}
\Big(\varepsilon_\mathrm{i}(z',\omega)
 \mathrm{Tr}[\overset{\text{\tiny$\leftrightarrow$}}{\mathbf{g}}_{\mathrm{ee}}^\dag(z,K,\omega,z')\cdot
 \overset{\text{\tiny$\leftrightarrow$}}{\mathbf{g}}_{\mathrm{ee}}(z,K,\omega,z')]\nonumber\\
&\hspace{0.5cm}+\mu_\mathrm{i}(z',\omega)
 \mathrm{Tr}[\overset{\text{\tiny$\leftrightarrow$}}{\mathbf{g}}_{\mathrm{em}}^\dag(z,K,\omega,z')\cdot
 \overset{\text{\tiny$\leftrightarrow$}}{\mathbf{g}}_{\mathrm{em}}(z,K,\omega,z')]\Big)\nonumber\\
&=\frac{\omega^3}{2\pi^3c^4}\sum_{\alpha,\beta}
\Big(\varepsilon_\mathrm{i}(z',\omega)
 |g_{\mathrm{ee}}^{\alpha\beta}(z,K,\omega,z')|^2\nonumber\\
 &\hspace{0.5cm}+\mu_\mathrm{i}(z',\omega)
 |g_{\mathrm{em}}^{\alpha\beta}(z,K,\omega,z')|^2\Big),
\label{eq:enldosstrat}
\end{align}
where $g_\mathrm{ee}^{\alpha\beta}$ and $g_\mathrm{em}^{\alpha\beta}$,
with $\alpha,\beta\in\{1,2,3\}$,
are components of the matrix representations of the spectral dyadic Green's functions
$\overset{\text{\tiny$\leftrightarrow$}}{\mathbf{g}}_{\mathrm{ee}}$ and
$\overset{\text{\tiny$\leftrightarrow$}}{\mathbf{g}}_{\mathrm{em}}$.
The NLDOS of the magnetic field is given by
\begin{align}
 &\rho_\mathrm{NL,m}(z,K,\omega,z')\nonumber\\
 &=\frac{\omega^3}{2\pi^3c^4}
\Big(\varepsilon_\mathrm{i}(z',\omega)
 \mathrm{Tr}[\overset{\text{\tiny$\leftrightarrow$}}{\mathbf{g}}_{\mathrm{me}}^\dag(z,K,\omega,z')\cdot
 \overset{\text{\tiny$\leftrightarrow$}}{\mathbf{g}}_{\mathrm{me}}(z,K,\omega,z')]\nonumber\\
&\hspace{0.5cm}+\mu_\mathrm{i}(z',\omega)
 \mathrm{Tr}[\overset{\text{\tiny$\leftrightarrow$}}{\mathbf{g}}_{\mathrm{mm}}^\dag(z,K,\omega,z')\cdot
 \overset{\text{\tiny$\leftrightarrow$}}{\mathbf{g}}_{\mathrm{mm}}(z,K,\omega,z')]\Big)\nonumber\\
&=\frac{\omega^3}{2\pi^3c^4}\sum_{\alpha,\beta}
\Big(\varepsilon_\mathrm{i}(z',\omega)
 |g_{\mathrm{me}}^{\alpha\beta}(z,K,\omega,z')|^2\nonumber\\
 &\hspace{0.5cm}+\mu_\mathrm{i}(z',\omega)
 |g_{\mathrm{mm}}^{\alpha\beta}(z,K,\omega,z')|^2\Big).
\label{eq:hnldosstrat}
\end{align}

\subsection{Local densities of states}

As integrals of the electric and magnetic NLDOSs in Eqs.~\eqref{eq:enldosstrat} and \eqref{eq:hnldosstrat},
the electric and magnetic LDOSs are given by
\begin{align}
 \rho_\mathrm{e}(z,K,\omega) &=\frac{\omega}{2\pi^3 c^2}\mathrm{Im}\Big[g_\mathrm{ee}^{11}+g_\mathrm{ee}^{22}+\frac{\varepsilon(z,\omega)^2}{|\varepsilon(z,\omega)|^2}g_\mathrm{ee}^{33}\Big]
 \label{eq:eldos},\\
 \rho_\mathrm{m}(z,K,\omega) &=\frac{\omega}{2\pi^3 c^2}\mathrm{Im}\Big[g_\mathrm{mm}^{11}+g_\mathrm{mm}^{22}+\frac{\mu(z,\omega)^2}{|\mu(z,\omega)|^2}g_\mathrm{mm}^{33}\Big]
 \label{eq:hldos}.
\end{align}

\subsection{Interference density of states}

Using Eq.~\eqref{eq:apxpoynting} with $\varepsilon(\mathbf{r}',\omega)=\varepsilon(z',\omega)$,
$\mu(\mathbf{r}',\omega)=\mu(z',\omega)$,
$\langle\hat\eta(\mathbf{r}',\omega)\rangle=\langle\hat\eta(z',\omega)\rangle$, and
\begin{align}
 &\int\mathrm{Tr}[\overset{\text{\tiny$\leftrightarrow$}}{\mathbf{G}}_\mathrm{ee}(\mathbf{r},\omega,\mathbf{r}')\times
 \overset{\text{\tiny$\leftrightarrow$}}{\mathbf{G}}_\mathrm{me}^\dag(\mathbf{r},\omega,\mathbf{r}')]d^2R'\nonumber\\
 &=\frac{1}{4\pi^2}
 \int\hat{\mathbf{z}}\hat{\mathbf{z}}\cdot\mathrm{Tr}[\overset{\text{\tiny$\leftrightarrow$}}{\mathbf{g}}_{\mathrm{ee}}(z,K,\omega,z')\times
 \overset{\text{\tiny$\leftrightarrow$}}{\mathbf{g}}_{\mathrm{me}}^\dag(z,K,\omega,z')]d^2K,
\end{align}
\begin{align}
 &\int\mathrm{Tr}[\overset{\text{\tiny$\leftrightarrow$}}{\mathbf{G}}_\mathrm{mm}(\mathbf{r},\omega,\mathbf{r}')\times
 \overset{\text{\tiny$\leftrightarrow$}}{\mathbf{G}}_\mathrm{em}^\dag(\mathbf{r},\omega,\mathbf{r}')]d^2R'\nonumber\\
 &=\frac{1}{4\pi^2}
 \int\hat{\mathbf{z}}\hat{\mathbf{z}}\cdot\mathrm{Tr}[\overset{\text{\tiny$\leftrightarrow$}}{\mathbf{g}}_{\mathrm{mm}}(z,K,\omega,z')\times
 \overset{\text{\tiny$\leftrightarrow$}}{\mathbf{g}}_{\mathrm{em}}^\dag(z,K,\omega,z')]d^2K
\end{align}
gives
\begin{align}
 &\langle\hat{\mathbf{S}}(\mathbf{r},t)\rangle\nonumber\\
 &=\int\int_0^\infty
 \int_{-\infty}^\infty\frac{\hbar\omega^4}{2\pi^3c^3}\Big(\mu_\mathrm{i}(z',\omega)\nonumber\\
 &\hspace{0.5cm}\times\mathrm{Im}\Big[\hat{\mathbf{z}}\hat{\mathbf{z}}\cdot\mathrm{Tr}[\overset{\text{\tiny$\leftrightarrow$}}{\mathbf{g}}_\mathrm{mm}(z,K,\omega,z')\times
 \overset{\text{\tiny$\leftrightarrow$}}{\mathbf{g}}_\mathrm{em}^\dag(z,K,\omega,z')]\Big]\nonumber\\
 &\hspace{0.4cm}-\varepsilon_\mathrm{i}(z',\omega)\nonumber\\
 &\hspace{0.5cm}\times\mathrm{Im}\Big[\hat{\mathbf{z}}\hat{\mathbf{z}}\cdot\mathrm{Tr}[\overset{\text{\tiny$\leftrightarrow$}}{\mathbf{g}}_\mathrm{ee}(z,K,\omega,z')\times
 \overset{\text{\tiny$\leftrightarrow$}}{\mathbf{g}}_\mathrm{me}^\dag(z,K,\omega,z')]\Big]\Big)\nonumber\\
 &\hspace{0.5cm}\times\langle\hat\eta(z',\omega)\rangle dz'd\omega d^2K.
\end{align}
Note that the Poynting vector points purely in the $z$ direction, which is natural
due to the symmetry with respect to the $z$ axis.
Hence, the IFDOS can be written as
$\boldsymbol{\rho}_\mathrm{IF}(z,K,\omega,z')=\hat{\mathbf{z}}\rho_\mathrm{IF}(z,K,\omega,z')$,
where the scalar IFDOS $\rho_\mathrm{IF}(z,K,\omega,z')$ is given by
\begin{align}
 &\rho_\mathrm{IF}(z,K,\omega,z')\nonumber\\
 &=
 \frac{\omega^3n_\mathrm{r}(z,\omega)}{2\pi^3c^4}\Big(\mu_\mathrm{i}(z',\omega)\nonumber\\
 &\hspace{0.5cm}\times\mathrm{Im}\Big[\hat{\mathbf{z}}\cdot\mathrm{Tr}[\overset{\text{\tiny$\leftrightarrow$}}{\mathbf{g}}_\mathrm{mm}(z,K,\omega,z')\times
 \overset{\text{\tiny$\leftrightarrow$}}{\mathbf{g}}_\mathrm{em}^\dag(z,K,\omega,z')]\Big]\nonumber\\
 &\hspace{0.5cm}-\varepsilon_\mathrm{i}(z',\omega)\nonumber\\
 &\hspace{0.5cm}\times\mathrm{Im}\Big[\hat{\mathbf{z}}\cdot\mathrm{Tr}[\overset{\text{\tiny$\leftrightarrow$}}{\mathbf{g}}_\mathrm{ee}(z,K,\omega,z')\times
 \overset{\text{\tiny$\leftrightarrow$}}{\mathbf{g}}_\mathrm{me}^\dag(z,K,\omega,z')]\Big]
 \Big)\nonumber\\
 &=\frac{\omega^3n_\mathrm{r}(z,\omega)}{2\pi^3c^4}\nonumber\\
 &\hspace{0.5cm}\times\Big(\mu_\mathrm{i}(z',\omega)
 \mathrm{Im}\Big[g_\mathrm{mm}^{11}g_\mathrm{em}^{21*}-g_\mathrm{mm}^{22}g_\mathrm{em}^{12*}-g_\mathrm{mm}^{23}g_\mathrm{em}^{13*}\Big]\nonumber\\
 &\hspace{0.5cm}-\varepsilon_\mathrm{i}(z',\omega)
 \mathrm{Im}\Big[g_\mathrm{ee}^{11}g_\mathrm{me}^{21*}-g_\mathrm{ee}^{22}g_\mathrm{me}^{12*}-g_\mathrm{ee}^{23}g_\mathrm{me}^{13*}\Big]\Big).
 \label{eq:ifdosstrat}
\end{align}

\end{document}